# Encyclopedia of Electrochemistry

Article title: Electrochemistry and Optical Microscopy

Frédéric Kanoufi
Université de Paris, ITODYS, 75006 Paris, France
frederic.kanoufi@u-paris.fr

## Abstract
[Electrochemistry exploits local current heterogeneities at various scales ranging from the micrometer to the nanometer. The last decade has witnessed unprecedented progress in the development of a wide range of electroanalytical techniques allowing to reveal and quantify such heterogeneity through multiscale and multifonctionnal operando probing of electrochemical processes. However most of these advanced electrochemical imaging techniques, employing scanning probes, suffer from either low imaging throughput or limited imaging size. In parallel, optical microscopies, which can image a wide field of view in a single snapshot, have made considerable progress in terms of sensitivity, resolution and implementation of detection modes. Optical microscopies are then mature enough to propose, with basic bench equipment, to probe in a non destructive way a wide range of optical (and therefore structural) properties of a material *in situ*, in real time: under operating conditions. They offer promising alternative strategies for quantitative high-resolution imaging of electrochemistry. The first sections recall the optical properties of materials and how they can be probed optically. They discuss fluorescence, Raman, surface plasmon resonance, scattering or refractive index. Then the different optical microscopes used to image electrochemical processes are examined along with some strategies to extract quantitative electrochemical information from optical images. Finally the last section reviews some examples of *in situ* imaging, at micro- to nanometer resolution, and quantification of electrochemical processes ranging from solution diffusion to the conversion of molecular interfaces or solids.]

## Keywords
[Optical microscopies, fluorescence, surface plasmon resonance, scattering, Raman, refractive index, single entity, imaging, electrode, electrochemical conversion]

# Main text
## [1] Introduction

Suppose a compound $Ox$ is reduced by an electrode by a transfer of $m$ electrons to a compound $Red$, with respective stoichiometric coefficients $v_o$ and $v_r$, according to:

$$v_o Ox + me \leftrightarrow v_r Red \quad (1)$$

The reaction occurs in an electrode potential, $E_{el}$, region that makes the electron transfer thermodynamically or kinetically favorable.

For systems controlled by thermodynamics, the electrode potential relates the concentration of the compounds at time, $t$, at the electrode surface ($C_{Ox}(el,t)$ and $C_{Red}(el,t)$) according to the Nersnt equation:

$$E_{el} = E^0 + \frac{\mathcal{R}T}{mF} \ln \frac{C_{Ox}^{v_o}(el,t)}{C_{Red}^{v_r}(el,t)} \quad (2)$$

with $E^0$ is the standard potential for the reaction, $\mathcal{R}$ is the gas constant, $T$ is the Kelvin temperature and $F$ the Faraday constant.

Besides, the current, $i_{el}$, flowing the electrode owing to this transformation obeys Faraday's law and therefore reflects the rate of consumption of $Ox$ or of production of $Red$ (in mol s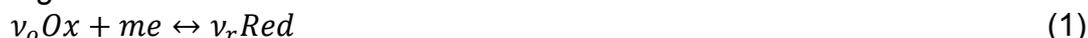⁻¹):

$$\text{Rate (mol/s)} = \frac{i_{el}}{mF} = -\frac{dN_{Ox}}{v_o dt} = \frac{dN_{Red}}{v_r dt} \quad (3)$$

where $N_{Ox}$ and $N_{Red}$ correspond respectively to the amount of moles of Ox and Red.

Using an appropriate electrode potential waveform, the rate of this transformation is inferred from the electrode current response, i.e. the current-potential, $i - E$, curve. This is fully described in electrochemical textbooks for various mechanistic configurations and potential or current waveforms [1].

At this point it is seen that $Ox$ consumption or $Red$ formation rates rely on the knowledge of the stoichiometric coefficients and of an ideal 100% Faradaic yield. If many mechanisms can be proposed from $i - E$ curves analysis, indirect and complementary titration of $Ox$ and/or $Red$ are usually required, and obviously such considerations apply equivalently to the understanding of non-faradaic contributions.

Many different analytical techniques are actively used to complement electroanalytical ones with *in situ* macroscale compositional and structural descriptions. Microscopes aim at probing these processes at the micron to nanometer scale. Electrochemical processes are inherently heterogeneous and quantifying or identifying such heterogeneity has fueled the growth of scanning electrochemical probe microscopies, SEPMs, over the last three decades. First popularized with ultramicroelectrodes by the scanning electrochemical microscopy, SECM, nowadays nanopipette probes offer routine sub-micron resolution imaging, in the scanning scanning electrochemical cell, SECCM, and ion conductance microsopies, SICM, configurations.

Despite their success, the main drawbacks of SEPMs are (i) a low imaging throughput (temporal resolution) and, (ii) a spatial resolution often difficult to control, as resulting from the intricate convolution of reaction/mass transfer within the probe-surface gap. Hence, the imaging resolution of most SPEMs studies is comparable to the optical diffraction limit. Morevoer, a whole field of view can be imaged in a single snapshot by optical microscopies, making them promising imaging alternatives for electrochemistry.

If the first reports on optical imaging of electrochemical processes date back to the SECM origin, in the late 1980s [2,3], it was not until the late 2000s, with the widespreading of ultra-sensitive charge-coupled device, CCD, cameras and the emergence of plasmonics, that optical microscopies have raised interest. The latter was further fueled with the recent super-localization concepts opening to nm-resolution imaging, overpassing the limits of diffraction, and the development of high-performance open-source image analysis tools, such as FIJI [4], which allow tracking dynamically motion events of a few nanometers associated with electrochemical processes across multiple individual entities on the same field of view.

If remarkable progress has been made with operando characterization techniques employing X-Ray, electron microscopy or neutron techniques, the advances made in optical microscopy together with the basic equipment needed, non-vaccum handling and its nondestructive nature makes the latter a promising *in situ/operando* monitoring of electrochemical processes. Hence, various optical microscopies have been coupled to most electroanalytical techniques to image most of the fields covered by electrochemistry (Figure 1). This chapter reviews the strategies developed, with particular attention to those imaging electrochemical processes *in situ* and in real time, at high spatial and temporal resolutions, and if possible quantitatively.

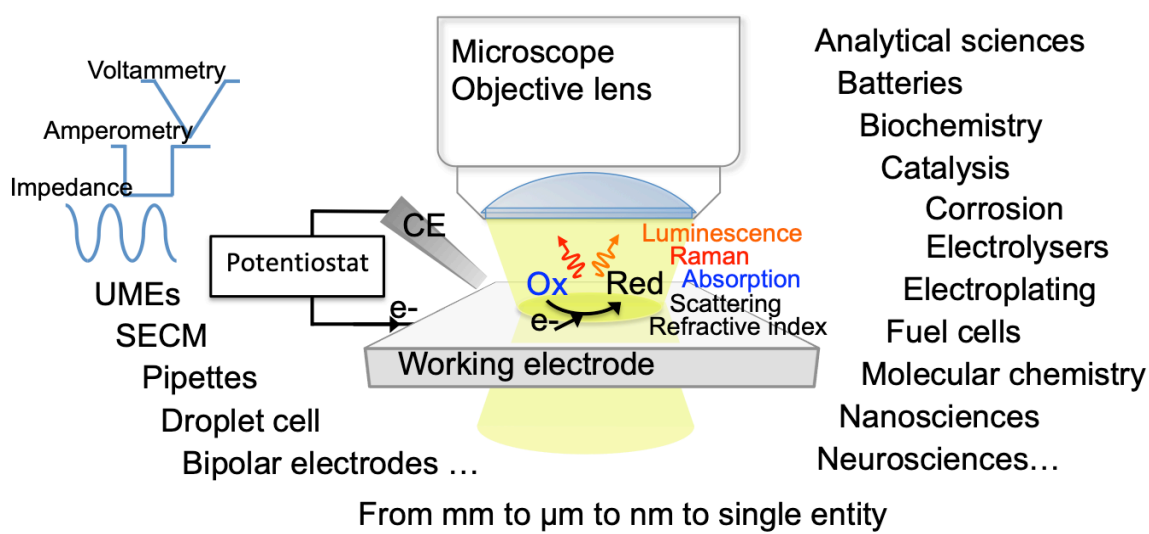

**Figure 1.** Examples of techniques and fields covered by optical microscopy monitoring of electrochemistry.

The most obvious situations concern the conversion of electrochromic compounds, which exploits a change in optical properties upon a change in redox state, meaning during (1). Electrochromism is usually related to a change in light absorption (change in color). Light absorption can be accompanied by a light emission process (luminescence) and compounds that exploit a change in light emission upon change in redox state (electrofluorochromism) are also prone to such optical microscopy inspections. It is also shown that optical microscopies are sensitive to changes in other fundamental material's optical properties, the index of refraction or the permittivity.

This section will first recall the optical properties of materials and the optical principles associated. The next three sections describe the optical microscopes used to monitor electrochemical processes and discuss the strategies employed to extract quantitative mechanistic information. The last section details selected examples where optical microscopies have been used to image and quantify *in situ* electrochemical processes, ranging from the diffusion of solution species to the conversion of molecular interfaces and solids.

# [2] Optical Principles

The purpose of this section is to give basic concepts for apprehending the different methods detailed in this chapter. It first recalls the characteristics of light as an electromagnetic radiation, then details the optical properties of a material and how these properties are revealed by observing its interaction with light.

## [2.1] Light: an electromagnetic radiation

Most of the principles outlined in the chapter can be understood considering the light as a propagating electromagnetic, EM, wave transporting energy through space. It is decomposed into an oscillating electric field, $\mathbb{E}$, and an oscillating magnectic field, $\mathbb{H}$, ($\mathbb{H}$ will not be discussed in the following). $\mathbb{E}$ and $\mathbb{H}$ are vectors perpendicular to each other and both are perpendicular to the propagation direction of the wave, described by a propagation wavevector $\mathbb{k}_\text{p}$ whose amplitude depends on the optical properties of the material under consideration.

A monochromatic light of **wavelength** $\lambda$ is characterized by the oscillation frequency of $\mathbb{E}$, $f = \frac{c}{\lambda}$ where $c$ is the celerity of light, or its angular **frequency**:

$$\omega = 2\pi f = \frac{2\pi c}{\lambda} \tag{4}$$

Meanwhile, for a plane wave, the orientation of $\mathbb{E}$ relative to the plane of incidence of the light beam describes its polarization (see Figure 2).

A linearly polarized plane wave can be described by a linear combination of p- or s-components, while an unpolarized light has equal contributions in each polarization.

The **energy** transported by the EM wave, is proportional the optical signal intensity, $I_{opt}$, given by an optical instrument, *e.g.* a microscope. Here $I_{opt}$ is the photon flux, or light intensity, $\Phi_0$ in photon/s, integrated during the image acquisition time, $\Delta t_{acq}$, of the instrument:

$$I_{opt} = \Phi_0 \Delta t_{acq} \tag{5}$$

and compared to an electrochemical charge. As $\Phi_0$ and $I_{opt}$ are proportional, only $I_{opt}$ isconsidered in the following. Since $\Phi_0$ is proportional to the square of the amplitude of the electric field $\mathbb{E}$, then $I_{opt} \sim |\mathbb{E}|^2$.

Light is also considered as a stream of photons, carrying individually a quantum of energy (in electronvolt, eV):

$$\mathrm{E}_{exc} = \frac{hc}{\lambda} \tag{6}$$

with $h$ the Planck constant.

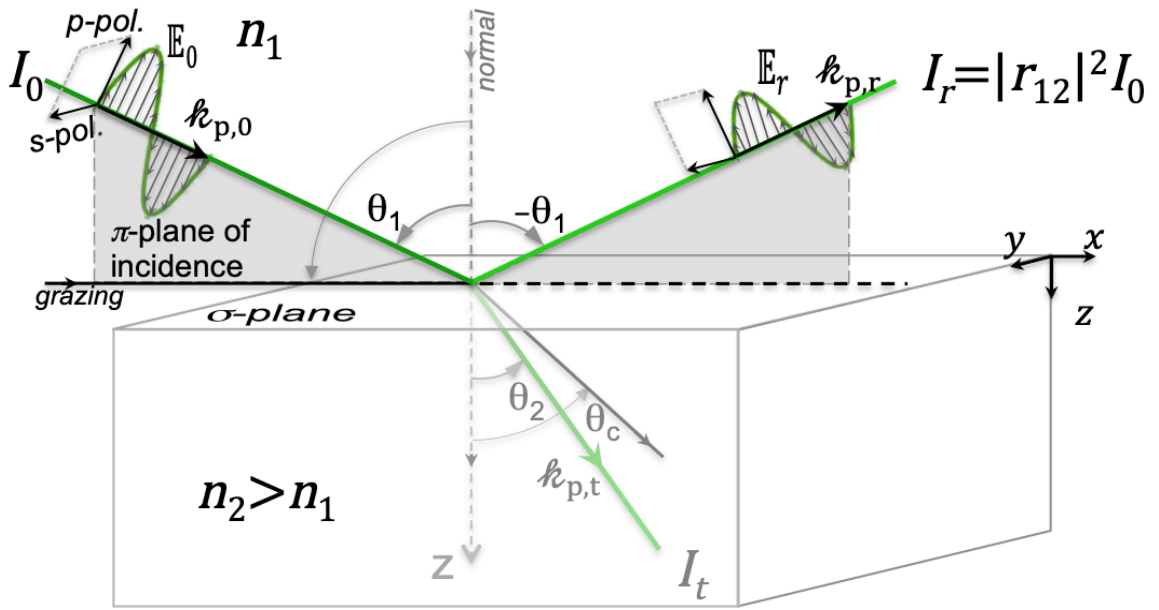

**Figure 2.** Reflection and transmission of an incident light beam of electric field $\mathbb{E}_0$ propagating in an optical medium ($n_1$) and impinging with oblique incidence (angle $\theta_1$ to normal) the interface with a different optical medium ($n_2$).

### [2.2] An optical medium and its optical and electromagnetic properties

The interaction of light with chemical matter is associated with macroscopic to microscopic descriptors. In geometric-optics an optical medium, of given chemical composition, is characterized by its index of refraction, $n$, and its absorption (or extinction) coefficient, $k$.

#### [2.2.1] Index of refraction, absorption

The **index of refraction** of a medium corresponds to the ratio of the celerity of light, $c$, to the rate of propagation of the wave, $v$, in this medium, $n = c/v$. The wavevector of the propagation of light in a medium $k_p$ is related to its propagation in vacuum, $k_{p0}$, and the refractive index of the medium,

$$k_\mathrm{p} = k_{\mathrm{p}0} n = \frac{2\pi}{\lambda} n = \frac{\omega}{c} n \tag{7}$$

For gases $n \approx 1$, for dielectric materials, such as liquid (electrolytes), transparent solids (glass, polymers,...), $n$ usually varies with the light wavelength, $\lambda$, and is comprised between 1.3 and 2, for water $n_w$=1.333.

$n$ is frequently employed to characterize organic compounds (known as $n^{20}_D$), or polymers molecular weight or size by light scattering techniques. Hence, $n$ is tabulated for many compounds [5,6] or can be estimated from additive molecular decomposition into functional groups [7]. Such decomposition is supported from the additive property of the molar refractivity, $R_M$ which is related to the index of refraction by Lorentz-Lorenz formula:

$$R_M = \frac{n^2-1}{n^2+2} V_M \tag{8}$$

with $V_M$ the molar volume. The molar refractivity, $R_{M,mix}$, of a mixture of two components, A and B, at molar fraction $x_A$ and $x_B$, with molar refractivities $R_{M,A}$ and $R_{M,B}$, then writes $R_{M,mix} = x_A R_{M,A} + x_B R_{M,B}$.

These different expressions allow predicting [8], from tabulated values, the index of refraction, $n_{sol}$, of a solution of a solute $S$ (*e.g.* salt of an electrolyte, up to 1 M [9]) at concentration $[S]$ in a solvent of $n_{solv}$, from:

$$n_{sol} = n_{solv} + \frac{R_{M,S}}{2}[S] \qquad (9)$$

$n$ and therefore the molar refractivity of materials usually depend on the light wavelength. This dependence is strongly marked for metals and semi-conductors. For dielectric materials, $n$ increases, by less than few percents, with decreasing wavelength ($n_{blue} > n_{red}$).

**Absorption coefficient.** An absorbing medium is able to absorb a portion of the light travelling through it, transforming it into another form of energy (heat or fluorescence). It is characterized by its absorption or extinction coefficient, $k$. Beer-Lambert law expresses the light transmitted, $I_t(z)$, in an absorbing medium as exponentially decaying with the travelled distance, $z$:

$$I_t(z) = I_0 e^{-\beta_A z} \qquad (10)$$

with $I_0$ the light flux entering the medium and $\beta_A$, in cm$^{-1}$, the power absorption coefficient or the inverse of the penetration depth of light, $\delta_A$:

$$\beta_A = \frac{4\pi k}{\lambda} = \frac{1}{\delta_A} \qquad (11)$$

In weakly absorbing materials ($k < 0.001$) light penetrates by $\delta_A > 80\lambda > 50\mu m$ ($\lambda$=633nm), for some polymers (polyvinylpyrolidone, $k \approx 0.005$) by ≈10μm, or in electrochromic iron or copper oxides $0.03 < k < 1$ by <2μm.

The constitutive molecular analogue of $k$ is the molar extinction coefficient, $\epsilon_M$ (in dm$^3$mol$^{-1}$cm$^{-1}$), which, for solutions of absorbing molecules, at concentration $[C]$, is obtained from $\beta_A = \epsilon_M[C] \ln 10$.

**Complex refractive index.** It is common accounting for both refraction and absorption in a medium using a complex number, called here the refractive index:

$$\tilde{n} = n + ik \qquad (12)$$

of real part $n$ and imaginary part $k$.

Different instruments are designed to evaluate $\tilde{n}$. For (organic) chemical compounds, liquids or solutions $n$ is measured using refractometers, while $k$ requires UV-vis spectrophotometers. For solids, metals, inorganic crystals, etc… complex refractive indices $\tilde{n}$ are determined by light reflection methods, such as ellipsometry and are often provided in suppliers libraries, or listed online [10].

### [2.2.2] Permittivity or dielectric constant

The refractive index is a macroscopic descriptor of the response of a material to a light beam. The permittivity, denoted $\varepsilon$, determines how much a material can polarize in response to an electric field, $\mathbb{E}$. This response consists in screening the electric field due to an internal rearrangement of charge (polarization). As $\tilde{n}$, the permittivity is a complex number $\tilde{\varepsilon} = \varepsilon_r + i\varepsilon_i$. The real part describes the material polarization (how much energy can be stored in the material). The imaginary part, the loss factor, evaluates how dissipative the material is to the EM field. It represents its electric conductivity.

Both optical and EM descriptors are related and $\tilde{\varepsilon} = \tilde{n}^2$ yielding $\text{Re}(\tilde{\varepsilon}) = \varepsilon_r = n^2 - k^2$ and $\text{Im}(\tilde{\varepsilon}) = \varepsilon_i = 2nk$, or $n^2 = \frac{\varepsilon_r}{2} + \frac{\sqrt{\varepsilon_r^2 + \varepsilon_i^2}}{2}$ and $k = \frac{\varepsilon_i}{2n}$.

Permittivity, as $\tilde{n}$, depends on the EM field frequency (wavelength). The real part $\varepsilon_r$ has the same physical significance as the (static or electronic) dielectric constant, $\varepsilon_S$ of the material, usually employed in electrochemistry for example to describe the electrical double layer effects. However $\varepsilon_S$ is a static "electronic" property determined at low frequency. The "optical" dielectric constant, $\varepsilon_i = \varepsilon_{op}$ rather describes the dynamic free carriers rearrangement upon polarization at the average high frequency $f = 5.45 \times 10^{14}$ Hz of the visible spectrum ($\lambda_{av} \approx 550$nm). Hence $\varepsilon_{op}$ is implied in the reorganization terms in Marcus and related electron transfer theories [11,12]. It ensues that $\varepsilon_S$ is larger than $\varepsilon_{op}$: for water at 25°C and 1atm $\varepsilon_{S,w} = 78.4$, while in the visible range $\varepsilon_{op,w} = n_w^2 = 1.77$. The difference is even larger when increasing the free-carriers density, in conductors. It should be noted that the imaginary part of $\varepsilon_{op}$ relates to absorption phenomena. The permittivity is then sensitive to electronic transitions which can be due to band gaps in semiconducting materials, or for metals to interband transition in the conduction band or transition from d-band to the conduction band. This is manifested in the spectral evolution of $\tilde{\varepsilon}$ (or of $\tilde{n}$) as resonance peaks.

The polarizability, $\alpha$, of a molecule or component of a material describes the effect of E field on such microscopic domains behaving as dipoles. It is related to ε in an additive constitutive relationship analogous to those of $\tilde{n}$ (Lorentz-Lorenz or Clausius-Mossotti). The (complex) polarizability, $\widetilde{\alpha_p}$, of an individual spherical entity (*e.g.* molecule, nanoparticle, NP) of permittivity $\tilde{\varepsilon}$ and volume $V_p$, embedded in a dielectric medium (*e.g.* electrolyte of permittivity $\varepsilon_m$) is then given by

$$\frac{\tilde{\varepsilon}-\varepsilon_m}{\tilde{\varepsilon}+2\varepsilon_m} = \frac{\widetilde{\alpha_p}}{3} V_p \varepsilon_m \qquad (13)$$

**[2.2.3] The special case of metals – surface plasmons**

At low frequencies, metals are electronic conductors, meaning they reflect EM waves. They depart from this ideal conductor behavior from the near IR frequencies where EM waves can be transmitted in metals up to tens of nm: a 50 nm thickness Au layers is a semi-transparent electrode. In the UV range, they behave as dielectrics, being transparent to light with some characteristic frequency absorption associated to electronic transitions. Apart from interband transition within the conduction band, CB, transitions from d bands to the CB explains the color gold or copper.

**Drude Model**

The interaction of EM with a metal is related to the motion of free carriers (establishment of an electrical current) upon its polarization. The Drude model describes it by expressing the motion dunamics of a free electron, considered as an individual non-interacting component of a gas (a plasma), and allowed to move within a fixed positively charged lattice (frozen at high frequency). When submitted to an oscillating electric field $\mathbb{E}$, the free electron displaces following the oscillations of $\mathbb{E}$, the immobile lattice acting as a restoring spring force (Figure 3). This results in a collective oscillation of the free electrons of the plasma, called plasmon. The motion dynamics reveals the natural frequency of the free oscillation of the plasma, or plasmon resonance, PR:

$$\omega_P = \frac{2\pi c}{\lambda_P} = \sqrt{\frac{N_e q_e^2}{m_e \varepsilon_0}} \qquad (14)$$

with the electron density $N_e$, unit mass, $m_e$, and charge, $q_e$. The resonance is typically in the UV range, *i.e.* 5-10 eV ($\lambda_P$ in the 240-120nm range). It ensues a theoretical expression for the metal permittivity:

$$\tilde{\varepsilon}(\omega) = 1 - \frac{\omega_P^2}{\omega(\omega - i\gamma)} \qquad (15)$$

where $\gamma$ characterizes the damping of the electron motion by collision. Such model is generally used to fit the spectral variations of experimental permittivities of metals and semiconductors. It can be extended by the Drude-Lorentz model, which considers high frequencies (beyond the spectrum of interest) permittivity, $\varepsilon_\infty$, and additional terms related to electronic transitions characterized by their resonance frequency, $\omega_{0,i}$, oscillator strength, $f_i$, and damping.

$$\tilde{\varepsilon}(\omega) = \varepsilon_\infty - \frac{\omega_P^2}{\omega(\omega - i\gamma)} - \sum \frac{f_i \omega_P^2}{\omega_{0,i}^2 - \omega^2 - i\gamma_i \omega} \quad (16)$$

Tabulated values of the plasma frequency and damping for different metals, together with example fittings can be found in [13].

These equations highlight, as a rule of thumb, the importance of PR in the optical properties of a metal. It particularly shows the sensitivity of PR to the electron density, $N_e$, on the metal, *i.e.* for electrochemistry, the electrode double layer charging. It also shows that interband electronic transition can stabilize the PR at a resonance frequency, $\omega_0$, lower than the natural plasmon resonance, $\omega_P$: *e.g.* the PR of Au is stabilized around its absorption wavelength (540nm).

Such model then allows depicting the conditions for EM wave excitation of the plasma of electron (photon-plasmon coupling). Light cannot excite volume plasmons but only surface plasmons, located at the interface between a metal and a dielectric. It results in a surface EM wave, evanescently confined, named surface plasmon polariton designated later on as surface plasmons SP. Figure 3 presents two types of SP modes that can be excited by light.

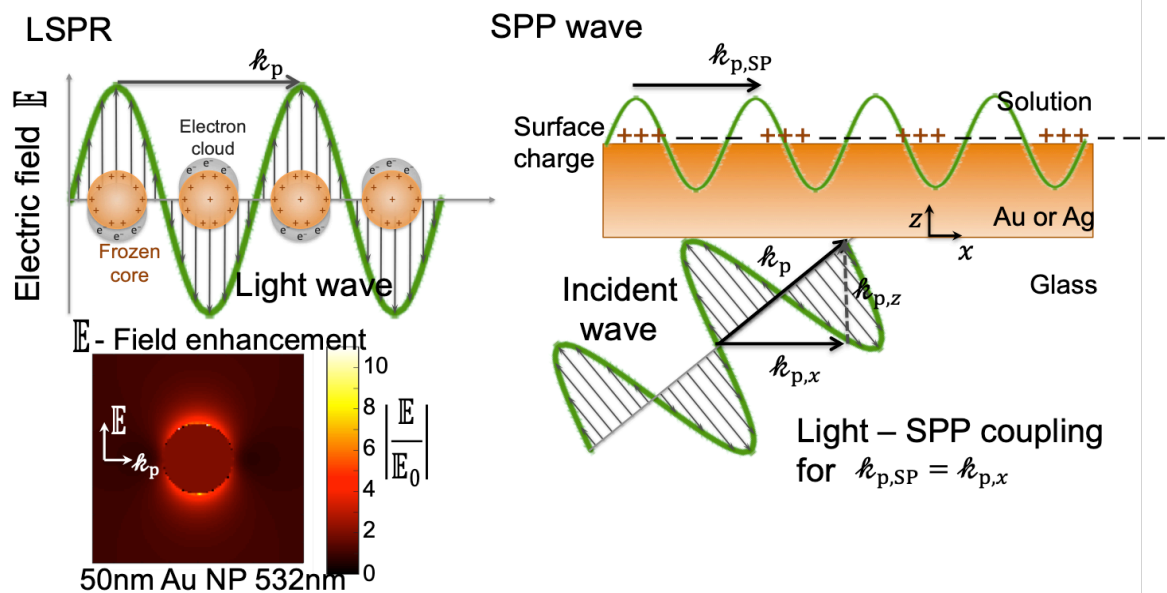

**Figure 3.** Generation of surface plasmons. (Left) Localized surface plasmon from the interaction of the electric field $\mathbb{E}$ of light with the free electrons of an atom/spherical NP (Drude model) and associated $\mathbb{E}$-field enhancement in the near-field of a 50nm Au NP illuminated by a 532nm light (with propagation and $\mathbb{E}$ directions). (Right) Conditions for producing propagating surface plasmon at surfaces (*e.g.* SPR microscopy).

**Localized surface plasmon**

In the first SP type, the SPs are confined in a metal NP of size smaller than the light wavelength. A localized SP can then be supported by the NP, whose resonance oscillation mode is reached for light excitation frequency $\omega_R$ given by:

$$\varepsilon_{r,NP}(\omega_R) + 2\varepsilon_m = 0 \qquad (17)$$

with $\varepsilon_{r,NP}$ the real part of the metal NP permittivity, given by the Drude Model, and $\varepsilon_m$ the surrounding medium permittivity. The resonance condition, explained later, results from the maximizing of the absorption of the light by the NP. Noteworthy, the NP, acting almost as a conductor at low frequencies, behaves abruptly, at $\omega_R$, as an electric dipole strongly absorbing (and scattering) this light wavelength, $\lambda_R$.
Simplifying (17), the localized SP resonance, LSPR, mode

$$\omega_R \approx \frac{\omega_P}{\sqrt{1+2\varepsilon_m}} \text{ or } \omega_R \approx \frac{\omega_P}{\sqrt{\varepsilon_\infty + 2\varepsilon_m}} \qquad (18)$$

shows upon nanoconfinement a lower frequency ($\omega_R$) than the natural resonance frequency ($\omega_P$): metal NPs absorb light of longer wavelength, $\lambda_R$ (redshift), than bulk metals. The LSPR modes supported and confined at nanostructures are means to concentrate, guide and therefore manipulate light at the nanoscale. Beyond promises in nanooptic devices, plasmonic NPs have opened new opportunities in chemical sciences such as the ability to confine an EM wave at the NP surface, to trigger, from LSPR, chemical reactions or to probe molecular recognition by enhanced light emission (Raman or fluorescence).[14,15]
Consequently, plasmonic NPs can be readily detected at $\lambda_R$ in the visible spectrum, using simple dark-field optical microscope, DFM.
The resonance condition shows that $\omega_R$ is sensitive to: (i) electron density, $N_e$, *i.e.* to NP surface charge, (ii) polarization of its local environment, an increase in $\varepsilon_m$ decreases $\omega_R$, redshifting the LSPR. This has opened a vivid branch in single NP electrochemistry, owing to the LSPR ability to probe at the single NP level subtle changes in bulk or surface NP composition.[16] As dictated by the expression of $\omega_R$, these changes are tracked spectroscopically from changes in the color or scattering intensity.

*Propagating Surface plasmons*

1D or 2D metallic nanostructures, respectively nanowires or ultrathin films, also support plasmonic modes upon light illumination; the surface plasmon polariton then propagates as a planar EM wave (Figure 3). As for the 0D NP structure, the metal nanostructures must be irradiated to generate SPR waves, which can propagate up to several micrometers (e.g. bouncing back and forth between the two ends of nanowires) in only one direction at the interface between the metallic structure and its surrounding medium. Enabling SPR waves at thin metal films requires certain illumination conditions schematized in Figure 3. First as the SPR wave is propagated parallel to the film, only p-polarized light ($\mathbb{E}$ parallel to the incident plane) can excite the SPR wave. This imposes that the $x$-component of the propagation wavevector of the EM wave, $\Bbbk_{p,x}$, equals that of the SP wave, $\Bbbk_{p,SP} = \Bbbk_{p,x}$.
Then for phase-matching, SPR can only be produced in three-layer configurations where the thin metallic film is sandwiched between two dielectrics with **different** index of refraction. The light, transmitted from the high-$n$ medium (usually glass, $n_g \approx 1.5$), enables the propagation of a SPR wave at the opposite interface, between the metal and, for electrochemistry, an electrolyte ($n_w \approx 1.33$). The former condition for SPR wave propagation is given by

$$\mathcal{k}_{\mathrm{p},SP} = \mathcal{k}_{\mathrm{p}0}\sqrt{\frac{\varepsilon_w \tilde{\varepsilon}(\omega)}{\varepsilon_w + \tilde{\varepsilon}(\omega)}} \qquad (19)$$

while that for its excitation from a light propagating in glass with incidence, $\theta_i$, from Fresnel laws is $\mathcal{k}_{\mathrm{p},SP} = \mathcal{k}_{\mathrm{p}0} n_g \sin\theta_i$. It ensues that the most efficient coupling between the high-$n$ and low-$n$ medium, requires oblique incidence illumination satisfying

$$n_g \sin\theta_i = \sqrt{\frac{\varepsilon_w \tilde{\varepsilon}(\omega)}{\varepsilon_w + \tilde{\varepsilon}(\omega)}} \qquad (20)$$

As a first approximation, this simplifies to $n_g \sin\theta_i \approx n_w$, which, as $n_g > n_w$ provides a condition for $\theta_i$, close to the critical total internal reflection, TIR, at the glass/water interface (Figure 4).

Consequently, an evanescent (exponentially decaying) electric field, is associated to the SPR wave, traveling from the plasmonic Au film for a short axial distance (~300nm) into the electrolyte. Besides, the coupling condition (20) shows a dependence on the refractive index of the electrolyte, $n_w$. Altogether, this explains that SPR is sensitive to local changes in refractive index occurring in the very thin electrolyte layers adjacent to the Au film. This has popularized SPR-based techniques and microscopy as a label-free biomolecular analytical tool.

### [2.3] Manifestation of light action on matter
#### [2.3.1] Reflectance/absorbance/refraction

As a first approximation, objects larger than the light wavelength, $\lambda$, (typically >10$\lambda$) are considered as infinite objects for the propagation of light. Their imaging by optical microscopy, except for their edge, is ruled by the propagation of light in infinitely large media.

The principles of reflection, refraction and transmission are recalled in Figure 2 in the case of an infinitely thick material, sample 2, covered by an electrolyte, 1. A light beam illuminates the sample and a microscope objective collects either the light transmitted by the sample or reflected by its surface.

An incident radiation travels in medium 1 with flux $I_0$, and impinges on the object, with an angle $\theta_1$. One portion of the radiation is reflected in 1 with inversion of direction and reflection of a portion of the energy with flux $I_r$. Another portion is transmitted, with flux $I_t$, moves into medium 2 and bends (is refracted). Refraction illustrates the difference at which the wave propagates in the two different media. The refracted angle $\theta_2$, given by Snell's law, $n_1 \sin\theta_1 = n_2 \sin\theta_2$, accounts for the shorter path explored by the beam, where it propagates more slowly.

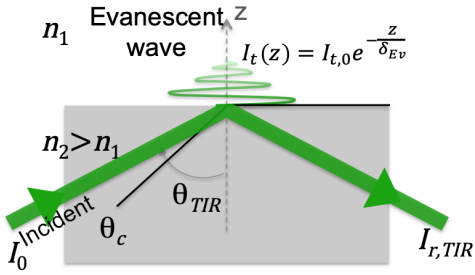
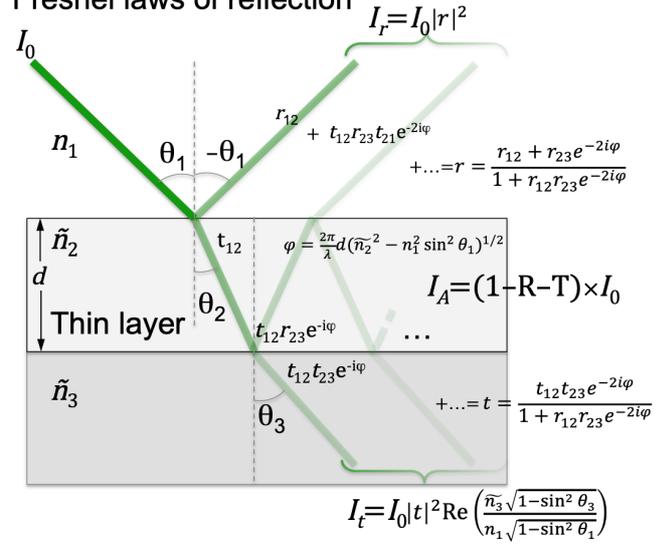

**Figure 4.** Schematic description of TIR illumination of an interface between two media ($n_2$, $n_1$), and of the Fresnel formalism at a two-interfaces between three media with reflection/absorption/transmission of a light beam, e.g. at a sample ($\widetilde{n_2}$) intercalated between an electrode ($\widetilde{n_3}$) and electrolyte ($n_1$).

### The case of Total Internal Reflection, TIR

Inversely a beam entering a medium of low-$n$ is refracted towards the interface, as schematized in Figures 2 and 4 by the limit or critical angle of observation,

$$\theta_c = \sin^{-1}\left(\frac{n_1}{n_2}\right), \text{ for } n_1 < n_2 \tag{21}$$

For incidence $\theta_2 < \theta_c$ the beam is partly transmitted in the low-$n$ medium. For $\theta_2 > \theta_c$ light is no more refracted in the low-$n$ medium and the incident beam is totally reflected by the interface, corresponding to TIR illumination (Figure 4). It is accompanied by the travelling of an evanescent wave along the interface with intensity that falls off exponentially with axial distance in the low-$n$ medium, characterized, as for absorption, by a penetration depth whose thickness $\delta_{Ev} = \frac{\lambda}{4\pi\sqrt{n_2^2 \sin^2(\theta_2) - n_1^2}}$ can be modulated with incidence angle in the 50-500nm range. This configuration then employs the evanescent wave to illuminate or excite optically objects present mostly in the thin penetration depth of medium 1.

This mode of detection is exploited in wide field optical microscopies such as TIR fluorescence microscopy (TIRFM) or SPRM to image processes within sub-wavelength region of an interface, or in scanning near-field optical microscopy to overcome the diffraction limit resolution.

### *Fresnel laws*

The conservation of light energy imposes that the flux of incident light is equal to the sum of the transmitted, reflected and absorbed fluxes. In this respect reflectance, $R = \frac{I_r}{I_i}$, transmittance, $T = \frac{I_t}{I_i}$, and absorptance, $A = \frac{I_a}{I_i}$, are defined as the ratio of their respective flux with respect to the incident flux, such as $R + T + A = 1$. Reflectance and transmittance are theoretically evaluated from Fresnel laws, and analytical expressions for different number or nature of interfaces and incident angle are given in [17]. They relate to the interfacial Fresnel coefficients $r = \mathbb{E}_r/\mathbb{E}_i$ respectively $t = \mathbb{E}_t/\mathbb{E}_i$ with, by conservation of the electric field, $t + r = 1$.
These coefficients, obtained from geometric considerations, depend on the p- or s-polarization of light. They are used in SPR or reflection-based microscopies. Noteworthy, the parameters, $\Psi$ and $\Delta$, devised in ellipsometric techniques express from the ratio $r_p/r_s = \tan\Psi \exp(i\Delta)$ [18].

The reflectance of an interface, obtained from the modulus of the field, yields in the simplest case of normal incidence:

$$R = |r|^2 = \left|\frac{n_1 - \widetilde{n_2}}{n_1 + \widetilde{n_2}}\right|^2 = \frac{(n_1 - n_2)^2 + k_2^2}{(n_1 + n_2)^2 + k_2^2} \quad (22)$$

It ensues estimates of the local refractive index variations of a material 2 or its environnement from absolute, $\Delta R$, or relative, $\Delta R/R$, reflectance variation. For more general cases, R and T are readily computed, or solved by online java applets [19].

In the case of thin layer, under electrochemical conditions local $\Delta R$ may be related to thin film deposition or conversion. The reflectance of a thin film, of refractive index $\widetilde{n_2}$ and thickness $d_l$, sandwiched between two optical media of respective refractive index $n_1$ and $\widetilde{n_3}$ is given from each interfacial Fresnel coefficient, as illustrated in Figure 4. This configuration describes many thin-film optical behaviors, from the enhanced sensitivity of anti-reflecting layers to the coupling of light with SPR.

Under normal illumination and in the limit of thin optical layers (i.e. $|\widetilde{n_2}|d_l < \lambda$) reflectance variation increases linearly with the layer thickness:

$$\frac{R_{layer} - R_{sub}}{R_{sub}} = \frac{\Delta R}{R} \approx 1 - \frac{8\pi}{\lambda} d_l n_1 \operatorname{Im}\left(\frac{\widetilde{n_2}^2 - \widetilde{n_3}^2}{n_1^2 - \widetilde{n_3}^2}\right) \quad (23)$$

where $R_{sub}$ and $R_{layer}$ are the reflectance of the bare or coated substrate respectively. More complex situations are given in [17].
Multilayers are treated alike through iterative consideration of the reflection/transmission at each layer [20].

### [2.3.2] Scattering

When a molecule or object is illuminated with a light beam, the photons can be transmitted (refraction), absorbed, or scattered, i.e. diffracted or reflected (Figure 5). In some cases absorption results in an electronic transition, through the formation of an excited state which can be accompanied by luminescence, a radiative deactivation (reemission of a photon of lower energy). The photon scattering can be either elastic without loss of energy or inelastic (Raman scattering).

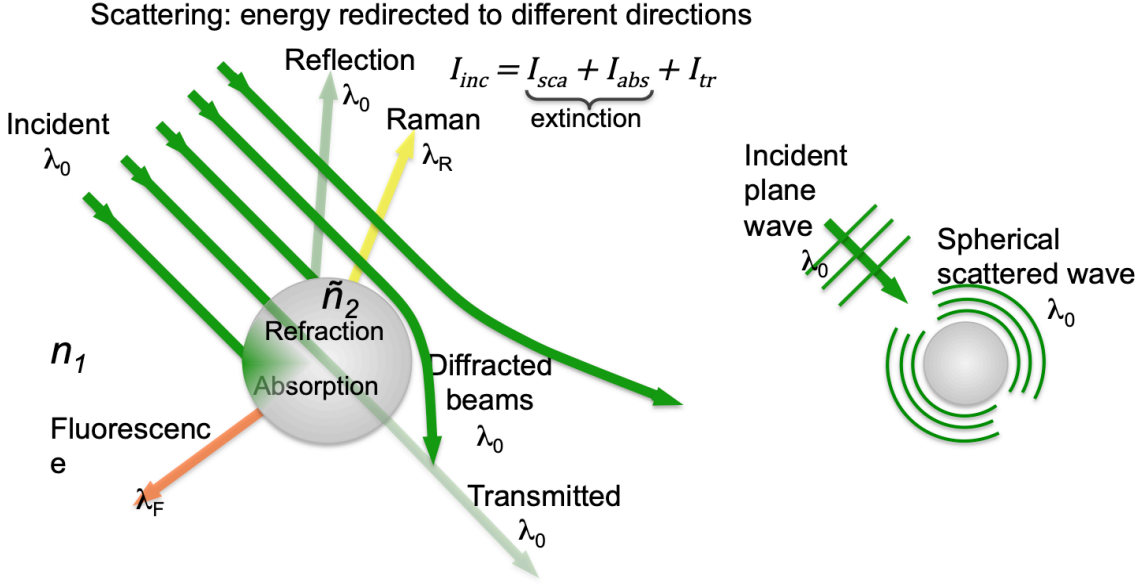

**Figure 5.** Schematic geometric-optics ray-approach of the different sources of interaction of light with a sphere: transmission and extinction by scattering and absorption with possible emission (fluorescence) or inelastic scattering (Raman). From the interaction of the incident plane wave with the sphere (see Figure 3), a resulting spherical scattered wave is produced.

The theory on molecular scattering developed by Lord Rayleigh has been extended by Mie for any size spherical particle. It evaluates the light flux redirected in different directions by a particle, *i.e.* the amount of light not directly transmitted (light extinction path). The scattering is characterized by a cross-section $\sigma_{sca}$, or efficiency $Q_{sca} = \frac{\sigma_{sca}}{\sigma_{geo}}$ when compared to the geometrical, $\sigma_{geo} = \frac{\pi}{4} d_p^2$, projected area of the particle of diameter $d_p$. Hence for an incident beam flux, $I_0$, impinging the particle, the scattered flux density is given by:

$$I_{sca} = \frac{\sigma_{sca}}{\sigma_{geo}} I_0 \tag{24}$$

The cross-sections for incident light absorption, $\sigma_{abs}$, and overall extinction, $\sigma_{ext} = \sigma_{abs} + \sigma_{sca}$, are similarly evaluated. They can be computed in various languages [21,22], or directly online [23–25]. Examples of scattering and absorption cross-sections of plasmonic, non plasmonic metals and dielectric NPs evaluated in water are compared in Table 1 to their geometrical one, revealing when $\sigma_{ext} > \sigma_{geom}$ the strong distortion of light (EM field enhancement) around metallic NPs [26], see Figure 3.

In the approximation of small spherical particle the first term of the expansion with respect to $d_p/\lambda$ for these cross-sections yield:

$$\sigma_{sca} \approx \frac{2\pi^5}{3} \frac{d_p^6}{\lambda^4} \varepsilon_m^2 \left|\frac{\widetilde{\varepsilon_{rel}}-1}{\widetilde{\varepsilon_{rel}}+2}\right|^2 + \cdots = \frac{8\pi^2}{\lambda^4} |\widetilde{\alpha_P}|^2 + \cdots \tag{25}$$

$$\sigma_{abs} \approx \frac{\pi^2 d_p^3}{\lambda} \varepsilon_m^{\frac{1}{2}} \text{Im}\left(\frac{\widetilde{\varepsilon_{rel}}-1}{\widetilde{\varepsilon_{rel}}+2}\right) + \cdots = \frac{3\pi^2 d_p^3}{\lambda} \varepsilon_m^{\frac{1}{2}} \left(\frac{\varepsilon''_{rel}}{\varepsilon_{rel}+2}\right) + \cdots$$

$$\approx \frac{2\pi^2}{\lambda \varepsilon_m^{1/2}} \text{Im}(\widetilde{\alpha_P}) + \cdots \tag{26}$$

where $\widetilde{\varepsilon_{rel}} = \frac{\widetilde{\varepsilon}}{\varepsilon_m}$ is the (complex) permittivity of the particle relative to its surrounding medium, and $\widetilde{\alpha_P}$ the particle polarizability.

| Nanoparticle Material / $d_p$ (nm) | $\sigma_{geo} = \frac{\pi}{4}d_p^{\,2}$ (nm²) | $\lambda_{m,abs}$ (nm) | $\dfrac{\sigma_{abs}}{\sigma_g}$ | $\lambda_{m,sca}$ (nm) | $\dfrac{\sigma_{sca}}{\sigma_g}$ | $\dfrac{\sigma_{ext}}{\sigma_g}$ |
|---|---|---|---|---|---|---|
| Au 15 | 1.8x10² | 520 | 0.99 | 524 | 0.004 | 0.99 |
| Au 100nm | 7.9x10³ | 590 | 2.6 | 536 | 4.5 | 7.0 |
| Ag 40nm[a] | 1.3x10³ | <400 | 0.25 | <400 | 0.55 | 0.8 |
| Ag 100nm | 7.9x10³ | 470 | 0.3 | 480 | 7.02 | 7.3 |
| Pt 40nm[a] | 1.3x10³ | <400 | 0.8 | <400 | 0.09 | 0.89 |
| Pt 100nm | 7.9x10³ | 476 | 1.8 | 430 | 2 | 3.7 |
| SiO₂ 40nm[a] | 1.2x10³ | - | 0 | <400 | 2.6x10⁻⁴ | 2.6x10⁻⁴ |
| SiO₂ 100nm[a] | 7.9x10³ | - | 0 | <400 | 8.4x10⁻³ | 8.4x10⁻³ |

**Table 1.** Geometrical cross-section, and absorption, scattering and total efficiencies of plasmonic Au, Ag and non-plasmonic Pt or dielectric SiO₂ NPs evaluated in water at $\lambda_m$ or 450nm[a].

Mie theory, developed for a spherical particle, has been extended to other spherical or cylindrical symmetry particles embedded in an homogeneous medium. It does not apply strictly to electrochemical situations considering particles on electrode (conducting) surfaces, or the multivarious geometries offered by colloid chemical synthesis or lithographic protocols.

Computational strategies are then employed to evaluate these scattering situations [27] under different electrochemical situations [28–31]. The volume integral discrete dipole approximation, DDA, method approximate a particle by a lattice of excited discrete dipoles, small compared to $\lambda$ [32,33]. Various codes are available open-source on repositories websites [33]. Boundary or finite element methods (Comsol) are more flexible in term of object geometries but rely on implicit functions for satisfying the different boundary conditions. The finite different time domain, FDTD, is another discretized method very popular in nanophotonics. However it is bound to a Cartesian mesh of the volume, which is demanding in computing time for accurate simulation of complex structures. Some programs with libraries employing these different methods are freely distributed [34–37].

Being aware of the limitations of models, Mie expressions depict, as a rule of thumb, the optical properties of NPs. Since $\sigma_{sca}$ increases as $\lambda^{-4}$ from (25) the shorter the wavelength the stronger the scattering. Besides, since from (25) $\sigma_{abs}$ increases as the particle volume, $V_P$, while from (26) $\sigma_{sca}$ increases as $V_P^{\,2}$, absorption dominates the optical response of small particles. Microscopies relying on the measurement of light absorption should be preferred to image small NPs to those relying on light scattering, such as DFM. Different alternatives have been proposed, one consists in using interference-based imaging [38] as in SPRM [39] to recover an optical response proportional to $V_P$.

Both (25) and (26) show a singularity when the denominator of the polarizability is minimized, i.e. when the real part of the particle permittivity $\mathrm{Re}(\tilde{\varepsilon}) = \varepsilon_r$ matches $-2\varepsilon_m$, a condition (18) fulfilled with metallic particles showing $\lambda$-dependent permittivity. This results in a maximum of the extinction cross-section.

An important application of Mie theory exploits the electromagnetic field enhancements in surface-enhanced Raman scattering (SERS).

*Raman scattering microscopies*

If in elastic scattering the energy of the photons is conserved, an inelastic diffusion, in a Raman scattering process, implies an exchange of energy between the photons and the molecules through the excitation of a molecular vibration or rotation mode. When the illuminated molecules absorb a part of the photons energy, the loss of energy of the scattered photons corresponds to Stokes scattering, and its wavelength is then redshifted. Alternatively, a gain of energy corresponds to anti-Stokes scattering and a blueshift of the wavelength.

The spectrum of Raman scattering then serves as a molecular fingerprinting of a medium. Unlike infrared, IR, absorption spectroscopy, which is therefore visible only in the IR spectrum, Raman scattering can be probed in all regions (UV, vis and IR) allowing local spectroscopic chemical identification of materials at ~0.5µm spatial resolution. However, the Raman scattering cross-section is several orders of magnitude lower than that of the Rayleigh (elastic) scattering, and Raman microscopy mostly images and chemically identifies >micron-sized objects, using highly intense laser illumination sources.

It is possible breaking both sensitivity and resolution limit by exploiting the increase of the Raman intensity with the fourth power of local EM field, which can be concentrated in very small volumes of space (hot spots), created within nanoscale metallic sharp edges or gaps. This is reached in the near field of optically excited plasmonic nanostructures, and one talks about plasmon-enhanced Raman scattering modes. By assembling metal-based NPs (even silica-coated metal core-shell NPs [40]), roughening metal surface or by lithography, a nanostructured surface is obtained which, upon laser illumination, is active for widefield imaging by SERS microscopy. In tip-enhanced Raman scattering, TERS, the metal tapered tip of a scanning probe microscope placed in the near-field of a metallic substrate and illuminated by a laser, also affords the enhancement required for the detection of the Raman signature of molecules present in the tip-substrate junction.

SERS and TERS provide images of surfaces with chemical identification at the single molecule level, respectively in a widefield and in a scanning observation mode. They are then mostly used to image the occurrence of chemical reactions through the disappearance (reactant) or appearance (product) of Raman peaks. So far, SERS is much easier to operate, shows higher enhancement factors and has shown much more advances than TERS. Indeed, apart from the delicate optical microscopy configurations [41], one challenge in TERS concerns the lack of reproducibility in the production of the tips. Moreover, the TERS hot spot is often altered during the time-consuming raster-scanning image acquisition.

### [2.3.3] Light Emission - Luminescence

Luminescence is a radiative process corresponding to the emission of a photon following a deactivation via an electronic transition of an excited state. The energy liberated by the photon emission may have been initially stored by different means: a chemical or electrochemical reaction (chemiluminescence or electrochemiluminescence, ECL, respectively), or the prior absorption of a photon of higher energy (photoluminescence, PL).

Figure 6 shows the most common energy level structure, or Jablonski diagram, of a molecular luminophore. It possesses an electronic singlet-singlet transition $S_1 \leftarrow S_0$ where an excited level of the first excited singlet-state, $S_1$, can be populated either by the absorption of a photon (a light beam) of sufficient energy, $h\nu$, in PL, or by a strongly exergonic chemical reaction, for CL or ECL. Fluorescence is the direct radiative return from $S_1$ to $S_0$, occuring 1-10ns after the formation of $S_1$.

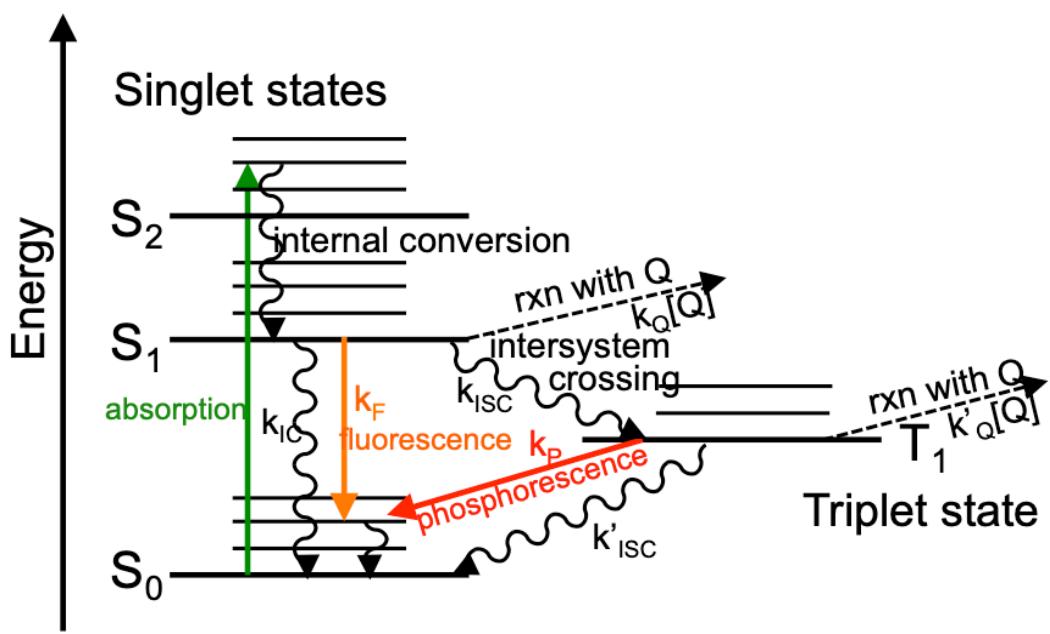

**Figure 6.** Typical energy level scheme (Jablonski diagram) explaining the absorption of a photon and reemission of a lower energy photon by a luminescence process.

Luminescence competes with different nonradiative deactivation routes such as: (i) chemical decomposition of the luminophore, (ii) chemical reaction with surrounding molecules (photobleaching or photoinduced electron transfer reactions), (iii) transfer of energy – by internal conversion and heat exchange with the solvent or surrounding molecules, *e.g.* by Förster resonant energy transfer (FRET) and, (iv) decay through a lower energy state, such as the triplet-state $T_1$. All these nonradiative processes occurring on purpose or involuntarily reduce the overall radiative rate of a luminophore, but in turns are means of studying their kinetics.

The radiative and overall non-radiative rates, $k_{rad}$ and $k_{nrad}$ respectively, give the luminescence quantum yield (equivalently emission probability):

$$\phi_L = \frac{k_{rad}}{k_{rad}+k_{nrad}} \quad (27)$$

Based on the Jablonski diagram considering only emission by fluorescence, the sum of the radiative and non radiative rates for the deactivation of the $S_1$ excited state describes its lifetime, $\tau_{S1} = \frac{1}{k_F+k_{IC}+k_{ISC}+k_Q[Q]}$ and:

$$\phi_F = k_F \tau_{S1} \quad (28)$$

Obviously, the quantum yield should be as close to 1 as possible. It depends on the presence or absence of a quencher $Q$ in solution (the larger $[Q]$ the smaller $\phi_F$). Comparing the quantum yield (*i.e.* fluorescence intensity) in the absence, $\phi_{F0}$, and in the presence, $\phi_F$, of $Q$ is the Stern-Volmer formalism picturing the sensitivity of a fluorophore to a quencher:

$$\frac{\phi_{F0}}{\phi_F} = 1 + k_Q \tau_{S10}[Q] \quad (29)$$

with $\tau_{S10}$ the unquenched fluorophore lifetime.

The intersystem crossing conversion ($S_1 \rightarrow T_1$) if energetically favorable, is a less likely spin-forbidden transition. However it is is facilitated in molecular systems presenting strong spin-orbit coupling, containing heavy-atoms, or possessing metal-to-ligand charge transfer transitions, MLCT. The return from the excited $T_1$ state to $S_0$ can also involve a (spin-forbidden) radiative deactivation called phosphorescence. The $T_1$ state is much more stable and phosphorescence occurs within >0.1μs. It ensues a lower yield of light emission, meaning lower detection sensitivity, but also ensures using these excited states for longer experiment durations. Triplet-state emitting luminophores are also more used as photoactivated catalysts in chemical processes: their slower quenching rates are accessible by pulsed LED instead of ps lasers.

Similarly, the phosphorescence quantum yield is obtained from $\tau_{T1} = \frac{1}{k_P + k'_{ISC} + k'_Q[Q]}$ lifetime:

$$\phi_P = k_P \tau_{T1} \eta_{T1} \qquad (30)$$

with $\eta_{T1} = \frac{k_{ISC}}{k_{ISC} + k_F + k_{IC}}$ the efficiency of population of the $T_1$ state upon (photo)generation of $S_1$.

**Quantifying the luminescence intensity**

This is pertinent to apprehend either quantitative imaging or single entity fluorescence studies. Luminescence efficiency relates to the molecular absorption cross-section $\sigma_{abs}$, the effective area of the molecule able to capture photons from the incident beam. In first approximation, $\sigma_{abs}$ is related to the molecular extinction coefficient.

Let $I_0$ and $I_t$ be respectively the photon fluxes of the incident and transmited light through a sample of thickness $b$ containing a fluorophore at (low) concentration $[C]$, from Beer-lambert law, (10) and (11):

$$I_t = I_0 10^{-\epsilon_M b[C]} \approx I_0 (1 - 2.303 \epsilon_M b[C] + \cdots) \qquad (31)$$

The luminescence intensity, $I_L$, is then proportional to the amount of light absorbed by the sample. and $I_L \sim I_t - I_0 \sim 2.303 \epsilon_M b[C] I_0$, yielding for a single molecule:

$$\sigma_{abs} \approx \frac{2.303 \epsilon_M}{\mathcal{N}_A} \qquad (32)$$

with $\mathcal{N}_A$ the Avogadro to be compared to its geometric cross-section, $\sigma_{geom} = \pi d_m^2$ with $d_m$ the molecule size. The higher $\sigma_{abs}$, the lower the number of required incident photons. It rules, for PL, the level of noise that can be reached to detect (single entity) fluorescent events with the highest sensitivity.

The ns to μs timescale of luminescence suggests that mostly PL operated in the ms to s range, or electrochemical triggering is steady-state compared to light emission. For PL, the steady-state luminescence intensity, $I_{L,SS}$, is given from the incident intensity (or photon flux, (5)), the absorption cross-section and luminescence quantum yield:

$$I_{L,SS} = \frac{\sigma_{abs} \phi_L}{\sigma_{geom}} I_0 \qquad (33)$$

This equation provides grounds for quantitative interpretation of luminescence experiments, or consideration of the best value of the incident photon flux to obtain an image with acceptable signal to noise ratio (SNR). In reality, it should be weighted from the efficiency of the full optical microscope, reminding that <1% of the emitted photons are efficiently collected by a microscope equipped with a CCD camera.

**Molecules**

A rich bank of fluorescent molecules or quantum dots, QDs, is available from biochemical studies. Electrochemical studies preferentially use electrofluoro-chromic [42,43] molecules or materials for which only one of the accessible redox states has luminescent properties. Indeed by a control of the redox state of the molecule the light emission can be turned on or off. Some of the photophysical and chemical properties are given in Table 2.

**Photon budget.**

The collection efficiency is of paramount importance in fluorescence microscopies. Indeed, if upon luminescence the electron returns to its ground state, making it ready for another excitation cycle, the fluorescent event is limited because of photobleaching that often results from the irreversible photodecomposition of the fluorophore. When designing an experiment it is therefore important to consider both luminescence and photobleaching, $\phi_b$, quantum yields. Photolysis studies [44] afford $\phi_b$ and the maximum number of photons emitted by a single fluorophore, typically $N_{max} \sim 10^4$-$10^8$.

A careful "photon budget" analysis [45] rules the tradeoff between illumination power or acquisition time and the limited number of images to capture. The number of image acquisitions that can be performed on a sample must be evaluated, especially when considering surface-bound fluorophores or single molecule detection.

| compound | $\lambda_{exc}$ (nm) | $\varepsilon_M$ (M$^{-1}$cm$^{-1}$) | $\sigma_{abs}$(10$^{-3}$ nm$^2$) | $\lambda_{em}$ (nm) | $\tau_L$ (ns) | $\phi_L$ | $N_{max}$ |
|---|---|---|---|---|---|---|---|
| fluorescein | 494 | 6 x 10$^4$ | 23 | 525 | 4 | 0.90 | 3 10$^4$ |
| Rhodamine 6G | 530 | 1.2 x 10$^5$ | 46 | 555 | 3.8 | 0.45 | 1.1 10$^6$ |
| Resorufin | 570 | 5.6 x 10$^4$ | 21 | 585 | 2.9 | 0.41 | n.d. |
| Ru(bpy)$_3^{2+}$ | 450 | 1.5 x 10$^4$ | 5.7 | 630 | 200 | 0.04 | >10$^5$ |

**Table 2.** Photophysic and photochemistry of examples fluorophores used in electrochemistry [44,45,270–272].

### [2.3.4] Electrochemiluminescence based microscopy

A means to improve the photon budget is to perform luminescence in the absence of external light illumination, *i.e.* by a chemical (CL) or electrochemical (ECL) triggering of luminescence. ECL imaging strategies are operated in near-dark conditions and circumvent most issues such as autofluorescence or saturation of the fluorophore absorption. It has popularized ECL as a quantitative optical readout for bioanalysis [46] and more recently as a promising optical imaging tool [47,48].

To apprehend the opportunities afforded by ECL microscopy, different reviews detail the mechanisms implied in ECL reactions [49]. Some of them are presented here, focusing on a muchstudied fluorophore: tris-2,2'-dipyridyl-ruthenium(II), Ru(bpy)$_3^{2+}$, denoted Ru$^{2+}$. Its luminescence proceeds from different reactions, but all involve a bimolecular ET reaction producing the $^3$MLCT excited triplet-state, Ru$^{2+*}$. The luminescence is issued from an exergonic reaction between a strong oxidant and a strong reducer, which are easily generated by electrochemistry. The first situation consists in polarizing alternatively in the anodic and cathodic direction an electrode placed in a Ru$^{2+}$ solution, to electrogenerate respectively Ru$^{3+}$ and Ru$^+$. In the electrode diffusion layer, these species encounter and annihilate:

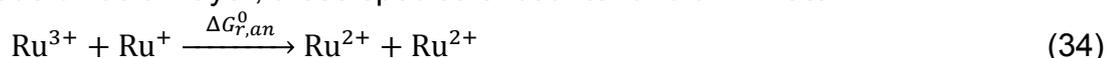
$$\text{Ru}^{3+} + \text{Ru}^+ \xrightarrow{\Delta G^0_{r,an}} \text{Ru}^{2+} + \text{Ru}^{2+} \tag{34}$$

The reaction is so exergonic ($\Delta G^0_{r,an} = F(E^0_{\text{Ru}^{2+}/\text{Ru}^+} - E^0_{\text{Ru}^{3+}/\text{Ru}^{2+}}) = -2.54$eV, see Figure 7A) that, in the realm of the Marcus inverted region, it demands an excess activation (slow kinetics). Instead, Ru$^{2+*}$ is produced by a less exergonic annihilation:

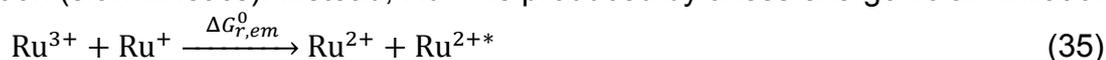
$$\text{Ru}^{3+} + \text{Ru}^+ \xrightarrow{\Delta G^0_{r,em}} \text{Ru}^{2+} + \text{Ru}^{2+*} \tag{35}$$

The standard potential of the $Ru^{3+}/Ru^{2+*}$ couple is approximated by $E^0_{Ru^{3+}/Ru^{2+*}} \approx E^0_{Ru^{3+}/Ru^{2+}} + \frac{hc}{\lambda_{em}}$, from which a slightly exergonic emission reaction ensues ($\Delta G^0_{r,em} = F(E^0_{Ru^{2+}/Ru^+} - E^0_{Ru^{3+}/Ru^{2+*}}) = -0.42$eV).

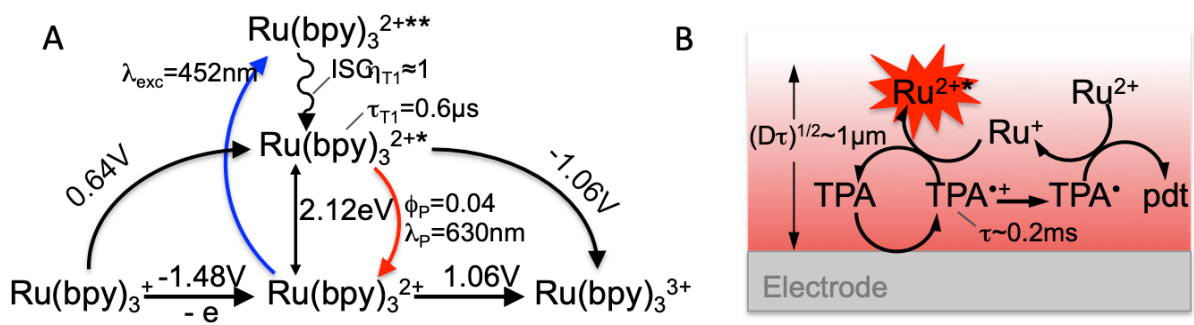

**Figure 7.** Principle of ECL. (A) energetics of the $Ru(bpy)_3^{2+}$ luminophore, and (B) chemically-limited lighting of thin layer of solution exploiting the limited lifetime of the electrogenerated intermediate radicals of a coreactant.

An alternative actuation mode of ECL, named "coreactant ECL", relies on a luminophore and a sacrificial coreactant. ECL emission is then stimulated by polarizing the electrode in a unique direction, to irreversibly transform the coreactant which, upon bond cleavage, produce a radical intermediate with opposite redox properties. The most popular configuration involves $Ru(bpy)_3^{2+}$ and an amine coreactant, such as tripropylamine, TPA. By oxidation, both $Ru^{3+}$ and TPA radical cation, $TPA^{•+}$, are formed, with, comparable oxidizing strengths ($E^0_{TPA^{°+}/TPA}$ ~0.9V) [50]. Upon dissociation, $TPA^{•+}$ yields the strongly reducing radical $TPrA^{•}$ able to reduce homogeneously $Ru^{2+}$ into $Ru^+$. Since near the electrode surface, different strong reducer and strong oxidant coexist, apart from the annihilation reaction (35) other reactions contribute to luminescence emission:

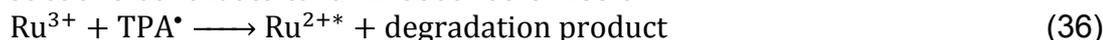
$Ru^{3+} + TPA^{•} \longrightarrow Ru^{2+*} + $ degradation product  (36)
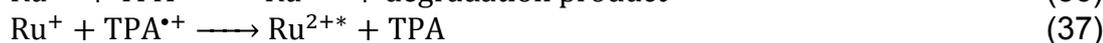
$Ru^+ + TPA^{•+} \longrightarrow Ru^{2+*} + TPA$  (37)

depending on the experimental conditions (luminophore and coreactant concentration, electrode material,...).

With a $\tau_{TPA^{•+}} \approx 0.2$ms lifetime, $TPA^{•+}$ is electrochemically generated in a $(D\tau_{TPA^{•+}})^{1/2} \approx 450$nm thin reaction layer adjacent to the electrode (Figure 7B). This chemical instability ensures an exponential decay of $TPA^{•+}$ concentration along the axial direction from the electrode, which, as in TIRF microscopy but without external illumination, allows triggering the luminescence of $Ru^{2+}$-labelled objects <1µm from an electrode [51].

The luminescence production is mostly steady-state, and the flux of photons, *i.e.* the luminescence intensity, $I_L$, is approximately ruled by the rate of $Ru^{2+*}$ production. For $Ru^{2+}$ this is obtained from the whole diffusion-reaction process modelling. For surface-immobilized $Ru^{2+}$, the $Ru^{2+*}$ production is dictated by the incoming flux of $TPA^{•}$. Full conversion of the generated $TPA^{•}$ into $Ru^{2+*}$ yields a maximal flux of emitted photon, and of luminescence, given by the electrochemical flux of $TPA^{•+}$ production, i.e. the TPA oxidation (by a 2-electron exchange), $i_{TPA}$, and the detector acquisition time, $\Delta t_{acq}$,

$$I_L \sim \frac{i_{\text{TPA}}}{2q_e} \Delta t_{acq} \qquad (38)$$

Typically a TPA oxidation current density ~2mA/cm$^2$ is at best equivalent to the emission from surface-tagged objects of ~10$^{15}$photon/cm$^2$/s or equivalent to the illumination of the object in PL with a light power density of $\frac{hc}{\lambda} \frac{\sigma_{geom}}{\sigma_{abs}\phi_L} \frac{i_{TPA}}{2} \approx 10$W/cm$^2$, *i.e.* low illumination power. Noteworthy, by controlling the oxidation current and the Ru$^{2+}$-label surface concentration, ECL should provide conditions close to single molecule fluorescence microscopy.

Obviously, a more rigorous analysis of the photon flux is obtained by FEM modeling of the reaction-transport processes. This was illustrated under different geometries: SECM [50], bipolar electrodes [52], micro- or nano-fabricated devices [53,54], or Ru$^{2+}$-labelled microbeads immobilized on electrodes [51].

## [3] Optical Microscope Configurations used in electrochemistry

Various optical techniques can characterize surface processes by different illumination and light collection modes (Figure 8).

In the first more conventionnal optical microscopes, the whole area of the surface of interest is illuminated and simulatenously imaged. The light, from a large field of view, is collected by a microscope objective and propagated to the 2D arrays detector of a CCD camera, and upon beamsplitting to a spectrograph, Sp, for spectroscopic imaging.

These techniques are subdivided into two subgroups depending on (i) if the surface and its environement are fully illuminated or (ii) if the illumination is confined.

The point scanning optical microscopies propose a sequential imaging system where the surface of interest is illuminated with the smallest possible spot and raster scanning it above the surface such that an image is constructed point by point.

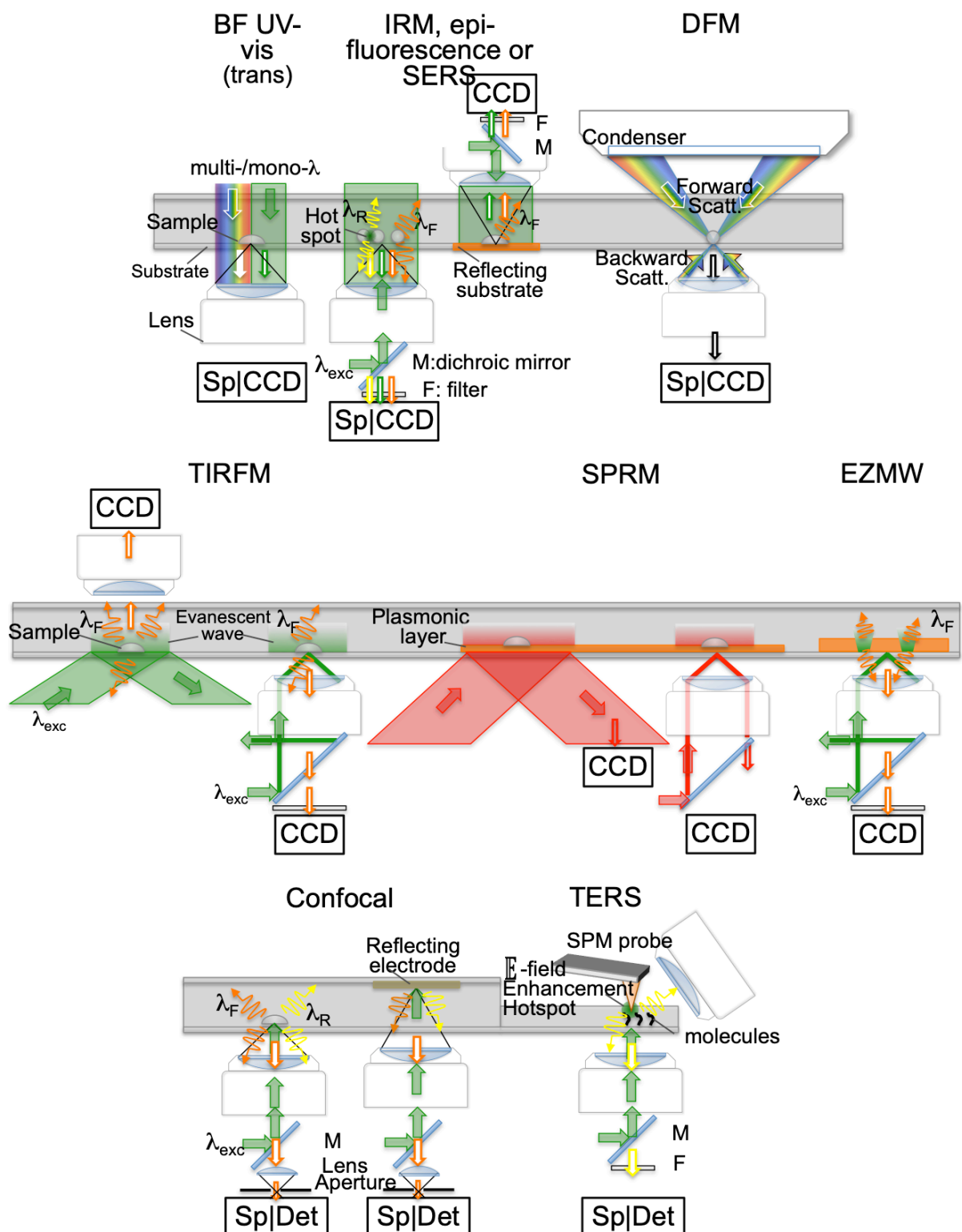

**Figure 8.** Illustration of configurations of optical microscopies to image the electrochemistry of a sample with an objective lens equipped with different detectors such as spectrograph (Sp) and a CCD camera or photomultiplier tube, PMT, detector (Det).

### [3.1] Microscope lens: magnification and collection efficiency

The signal received by a detector after passing through a microscope objective is first related to the magnification of the objective. Magnification is the portion of a field of view that is projected on a detector. A microscope objective of magnification $M\times$ projects a sample on a detector with a reduction of field of view by a factor $1/M^2$. The number of photons (or brightness) detected scales as $1/M^2$: a 100× objective collects only $N_M=100$ photons from a sample emitting $10^6$ photons per acquisition step (see Table 3).

However, the detection is entailed by noise from different sources: photon noise (shot or Poisson noise), dark current, pixilation, readout and background noise, all detailed in [55]. The photon noise arises from the stochastic nature of photon emission and Poisson counting statistics states that the noise is proportional to the squareroot of the signal. An increase in imaging magnification then results in a SNR decrease by $1/M$, from 1000:1 without magnification to 10:1 by 100×magnification.

To image with higher magnification without loosing too much photons (iso-brightness), high magnification microscope objectives, working at smaller distance from the sample as illustrated in Figure 9, should collect photons along a larger cone. This is the numerical aperture,

$$NA = n \sin \theta_{coll} \qquad (39)$$

defined from the half-angle of the lens collection cone, $\theta_{coll}$, and $n$ the index of refraction of the immersion medium (air, water or oil). Then the microscope objective collects within the collecting cone, a fraction $f_{NA}$ of photons emitted by an isotropic source:

$$f_{NA} = \frac{1-\cos\theta_{coll}}{2} = \frac{1-\sqrt{1-\left(\frac{NA}{n}\right)^2}}{2} \qquad (40)$$

For a same magnification, a 0.4 $NA$ air objective collects ~4% of emitted photons, while a 1.45 $NA$ oil immersion objective collect ~31% of them.
Table 3 compares the collection efficiency of some microscope objectives.

| Medium / $n$ | $M$[a] | $N_M$[b] | $SNR_M$[c] | $NA$[d] | $f_{NA}$[e] | $N_{obj}$[f] | $SNR_{obj}$[g] |
|---|---|---|---|---|---|---|---|
| Air / 1 | 5x | 40,000 | 200:1 | 0.15 | 0.012 | 480 | 22:1 |
| Air / 1 | 10x | 10,000 | 100:1 | 0.40 | 0.04 | 400 | 20:1 |
| Water / 1.33 | 60x | 278 | 16:1 | 1.30 | 0.39 | 109 | 10.4:1 |
| Oil / 1.51 | 100x | 100 | 10:1 | 1.40 | 0.31 | 31 | 5.5:1 |

**Table 3.** Collection efficiencies of microscope objectives. [a] magnification, [b] number of photons collected assuming $10^6$ photons are detected at 1x, [c,g] Poisson noise $SNR=N^{1/2}$, [d] numerical aperture, [e] fraction of photon collected from the $NA$ (no consideration of $M$), [f] resulting photons collected by the objective $N_{obj}=f_{NA}N_M$.

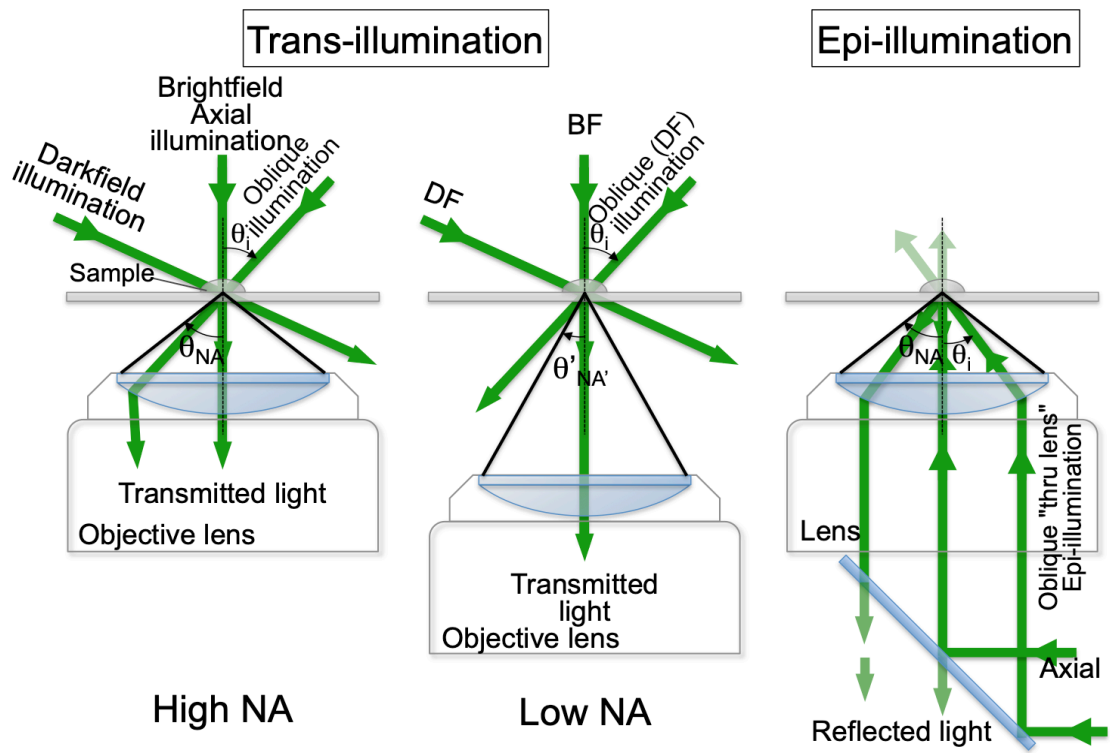

**Figure 9.** The numerical aperture, collection angle and mode of illumination/collection with a microscope objective.

### [3.2] Wide field microscopies
#### [3.2.1] Reflectance/absorbance based microscopies

The collection of light transmitted by a sample, e.g. deposited on a transparent electrode, ITO or thin Au-coated glass slide, is the simplest and oldest microscopic technique used, known as (transillumination) bright-field, BFM, microscopy, for imaging absorbance variations of a sample. The light reflected by the sample is collected from solution or electrode sides depending on the latter transparency under epi-illumination. Opaque or reflecting surfaces are imaged through the solution side, usually through thin layers and optical windows, requiring careful optical alignment to avoid stray light.

Interferometric detection is implemented in such microscopies. It is obtained by introducing a beamsplitter in the incident light path to produce interferences between a portion of the incident beam reflected on a reference mirror and the other one reflected on the surface, *e.g.* by dedicated Mireau objective.

Alternatively, in differential inference contrast, DIC, microscopy, a polarized light beam is separated into orthogonally polarized beams slightly physically separated and recombined (interfering) on the CCD detector. DIC images, apparently 3D, are sensitive to ~0.05 refractive index variations. DIC is often used in cell biology, not yet in electrochemistry, apart for a polarization contrast imaging of single NPs.

Besides, scattering nano-features in a reflection mode imaging also enables the production of intereference-patterned images. It defines a class of instruments named interference reflection microscopes, IRM, found with different acronyms in the literature (SP-IRIS [56], iSCAT [57], BALM [31,58], …). They produce images with interference fringes, owing to the mixing between the plane wave reflected by the solid/medium interface ($\mathbb{E}_r$) and the spherical wave scattered by nanofeatures ($\mathbb{E}_{sca}$), as in Figure 5. This detection mode is also exploited in SPRM. It affords enhanced NP visualization with sensitivities outperforming dark-field scattering observations. Indeed, the superposition of the two EM waves $\mathbb{E} = \mathbb{E}_r + \mathbb{E}_{sca}$ gives a light intensity,

$$I_{IRM} = |\mathbb{E}_r + \mathbb{E}_{sca}|^2 = I_r + I_{sca} + |\mathbb{E}_r||\mathbb{E}_{sca}|\cos\Theta \tag{41}$$

where $I_r = |\mathbb{E}_r|^2$, $I_{sca} = |\mathbb{E}_{sca}|^2$ are the intensities of the light reflected without scatterer, and scattered by the nanofeature respectively, and $\Theta$ refers to the phase angle difference between the fields.

According to Mie theory, $I_{sca}$ scales as $V_P^2$ while the last term scaling as $V_P$ becomes preponderant over $I_{sca}$ when the nanofeature size decreases. These techniques cover a wide range of label free imaging, from the characterization of thin films to single particle (>5nm) or even single protein detection [57]. They have been very recently introduced in electrochemistry.

### [3.2.2] Dark-field microscopy

Besides the axial trans- or epi-illumination modes, oblique illuminations at angle smaller than the half of the $NA$ of the microsope objective, are obtained by introducing the light beam in the extreme edge of the microscope objective (Figure 9).

Oblique illumination with angle larger than $NA/2$ of the microscope objective is named dark-field, DF, illumination, as neither the excited nor the directly transmitted or reflected light rays can be collected by the microscope objective. It creates a dark background from which only the light scattered by the excited sample in the direction of the objective is collected. It thus requires illumination at high angle and collection with a lower numerical aperture objective, which limits the number of photons collected and therefore the detection sensitivity. The NP size limit is ~40nm, for Ag, but can be improved to ~10nm when using a high $NA$ water immersion objective [59]. Commonly DFM is used upon trans-illumination, probing the forward scattered field. By restricting the field of view to a single NP, the scattered light collected by the microscope objective can be directed towards a CCD camera for imaging, and/or, *e.g.* after separation by a beam splitter, directed towards a spectrograph, Sp [29]. Color camera or hyperspectral dark field microscopes, e.g. via commercial platform (Cytoviva®, [60–63]) or hyperspectral camera, collect images with both full spectral identification and 2D localization of single NPs.

### [3.2.3] Epifluorescence microscopy

It is the simplest configuration used to image luminescent events. A laser or arc lamp creates a parallel (collimated) beam of light, which illuminates a sample through a microscope objective. For single molecule studies a laser beam is used with several tens of microns illumination area. The excited fluorophores within the focused light beam emit luminescence isotropically, and only the light emitted in the direction of the collection angle of the objective is collected. It passes through a dichroic mirror and a filter to eliminate residual excitation light before being directed to a CCD.

### [3.2.4] Superlocalization SERS microscopy

The epi-illumination configuration is also used in single molecule Raman detection [64]. It has been extended more recently for super-resolution Raman imaging of (few) molecules trapped within SERS hot spots [65,66].

To collect the Raman inelastic scattering, molecules must be excited by a stable laser beam with sufficient power (typically 100mW) to collect an image revealing the regions with Raman activity. The Raman light back-scattered towards the microscope objective is then split in two beams, one directed towards the CCD camera for acquiring the image, and a second part directed towards a spectrograph to obtain a Raman spectrum. Noteworthy if the signal collected at the CCD camera enables spatially resolved features, the Raman spectrum integrates the contribution of all Raman active features in the excited/imaged region. Single (or few) molecule imaging then requires working within a single SERS hot spot and with dilute (~nM or less) solution of the Raman active molecules, such that within the optically probed ~pL volume single (or few) molecule is present.

### [3.3] Imaging within confined space
#### [3.3.1] Imaging thin layers: total internal reflection microscopy

TIR microscopy employs the exponential decay of the evanescent field generated upon TIR at an interface between a high- and low-index media. For electrochemical situations this is usually obtained at the interface between an ITO electrode and an electrolyte, for angles of incidence larger than the critical angle. The advantage of TIR is that only the objects within the evanescent field of thickness $\delta_{Ev}$~100-500nm adjacent the electrode are illuminated. Compared with confocal microscopy, TIR provides 5 times thinner axial optical depth, leading to a dramatic reduction in background signal and increase in axial resolution. However, TIR is only useful for imaging processes within this thin layer of electrolyte. TIR fluorescence microscopy, TIRFM, is particularly used in single molecule studies owing to the considerably low volume probed, or for bioelectrochemical studies (cell membrane adhesion, exocytosis [67,68], bioassays…). Recent developments in TIR scattering or dark-field microscopies have demonstrated unprecedented sensitivity for the label free imaging of individual objects [69]. This should be fruitful for future electrochemical studies.

Two ways are commonly used to implement a TIR microscope.

In the first one (Kretschmann configuration), the transparent glass coated electrode is coupled to a prism with an index matching oil. Usually a laser beam hits the prism with an angle of incidence higher than the critical angle (TIR condition). The evanescent field illuminates objects within the penetration depth adjacent the electrode/electrolyte interface; their image is captured by a microscope objective.

In the second configuration (TIR 'through-the-objective'), the microscope objective illuminates under oblique incidence the interface and collects the image of the objects (Figure 9). For a typical glass surface ($n \approx 1.52$) contacted with an aqueous electrolyte ($n_w = 1.33$), the critical incidence angle θ$_c$=61° ($n \sin \theta_c = n_w$) requires a minimal $NA_{TIR} = n \sin \theta_c \geq 1.33$ for the objective. It is then mandatory to use a high $NA$ objective with high magnification (60x to 100x), then operating at low observation distance (<1mm) and contacting the glass substrate with an index-matching immersion oil.

SPRM is operated similarly in both configurations'. The former prism configuration is popular in bioanalytic assays for its >mm wide-field. 'Through-the-objective' illumination requires high $NA$, high magnification, objective with incidence angle ~70°, analogous to TIRFM, but here illuminated with Red to near-IR lasers.

#### [3.3.2] Zero-mode waveguides

Electrochemical zero-mode waveguides (E-ZMW) are conical nanopores in a thin metal film (forming a nanoelectrode) on an insulating optical window. The sub-wavelength size of the bottom of the nanopore provides a sub-wavelength aperture which confines optical irradiation and probing within the nanopore (with evanescent axial wave). The zeptoL volume of the pore provides a single molecule capture and analysis of both electrochemical and fluorescent processes [70–74].

### [3.4] Scanning microscopies,
#### [3.4.1] Confocal Microscopies

Different types of confocal microscopes are found, the most common ones are the confocal laser scanning microscope, CLSM, or the spinning disk confocal microscope.

In CLSM, a laser beam is reflected from a dichroic beamsplitter towards a high NA microscope objective such that the laser is focused to diffraction-limited spot at the region of interest. The higher the NA and the smaller the diffraction limited spot, the lower the residual background scattering.

For fluorescence excitation by CLSM, the emitted fluorescence is collected by the same microscope objective and goes through the dichroic beamsplitter. It is then focused by an additional lens through a pinhole aperture located from the microscope image plane, close to the detector. The pinhole rejects out-of-focus light in the image of the sample. Its diameter determines the depth of field of the confocal image. If the axial resolution is improved by closing partially the pinhole, it also severely limits the number of photons collected from the focal plane (reducing the $SNR$). It is often a misconception that CLSM is more appropriate in terms of resolution than widefield microscopy, besides it can be as little as 0.5% as efficient as wide-field systems.

CLSM is alo used to image light reflectance by a surface, and to generate Raman microscopy images (confocal Raman microscopy). The light collected is sent to a spectrograph, or for faster imaging rate to a hyperspectral camera, such that a Raman spectrum is collected at each scanned point.

#### [3.4.2] Tip enhanced microscopy

Despite the super-resolution localization of SERS reaction, the ultimate imaging resolution is reached in TERS by combining SERS specificity and enhanced sensitivity and the nanometer resolution capability of SPM tip coated by a plasmonic metal exploited for both nanoscale topography and Raman mapping.

In TERS, a laser light is focused onto a plasmonic metal tapered tip, enabling plasmon-enhanced Raman scattering conditions. This generates a local high electric field, a hot spot, in the region of the tip apex, enabling a ~$10^6$ local enhancement of the Raman signal. The illuminated tip is raster scanned over the sample (or the sample is moved). Thanks to the SPM force-feedback imaging, the tip provides a topographical mapping of the sample, the illuminated hotspot at its apex allows recording simultaneously the Raman photons inelastically scattered in its near field, enabling a nanoscale compositional mapping.

# [4] Retrieving Data
## [4.1] Image Analysis and Data Treatment

Driven by cell biological researches, fluorescence microscopies are now used as quantitative tools. This means characterizing an event (*i.e.* time) or an object (*i.e.* size) of interest with numbers which are most often represented by an optical intensity associated with spatial or temporal measurements.

While the pixel size determines the 2D resolution of a digital image, in confocal imaging, the spacing between *z*-stacks provides axial resolution. The important features of temporal measurement are often the acquisition time and time-lapse between acquisitions and also, when photobleaching occurs, the time duration of the experiment.

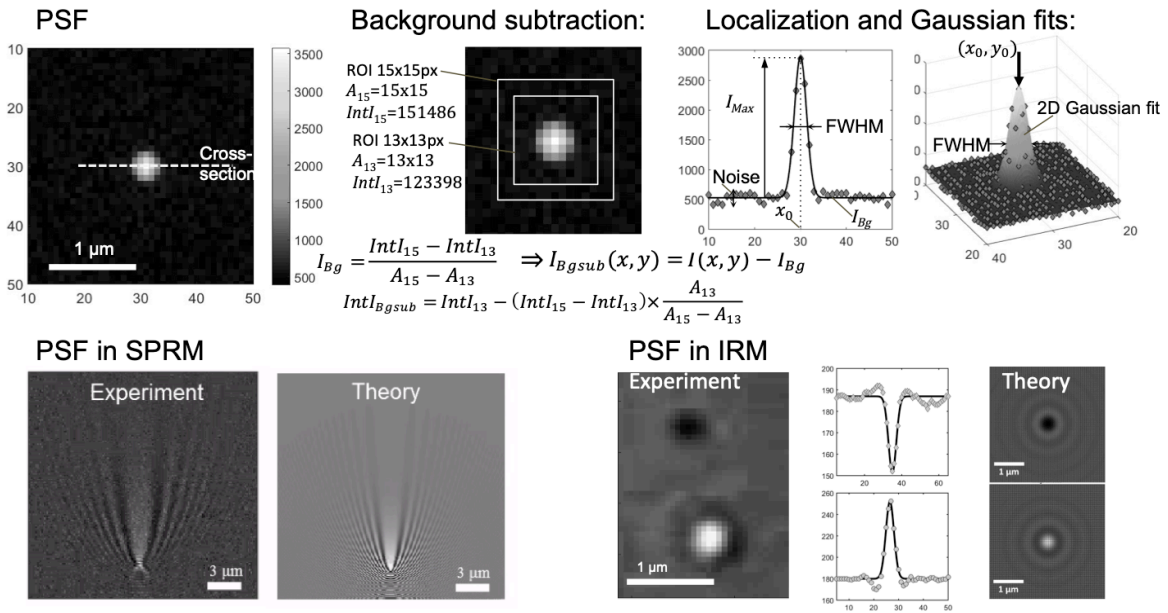

**Figure 10.** (Top) Extracting quantitative information from optical images: background, noise, localization, 1D or 2D PSFs. (Bottom) Interference-like PSF: large parabolic or ring-disks are predicted and observed in SPRM (adapted with permission from [96]) or IRM respectively; a Gaussian fit applies for the latter.

### [4.1.1] Considering the background

The information gathered from an optical microscopy experiment relies on images: intensity values of pixels are measured in a digital image to reveal the localization and quantity of light reflected/absorbed/emitted by a sample.

Background has various instrumental or physical origins. In reflectivity measurements (including SPR) it is the inherent reflectivity of the inert sample. In DF or fluorescence microscopies it is the scattering or fluorescence emitted from out-of-focus objects or from reflections in the optical cell...

Experimentally, the intensity of the background, $I_{Bg}$, surrounding an object in a region of interest, ROI, is obtained as explained in Figure 10 from the total integrated intensity, $IntI$, of two ROIs of slightly different areas, $A$. The signal of interest, $I_{Bgsub}(x,y)$ at each pixel of coordinates $(x,y)$, or its integrated intensity $IntI_{Bgsub}$, are obtained by subtracting the background contribution.

When dynamic processed is revealed from a series of images $Im^{1 \leq m \leq n}$, the first image, $Im^1$, can be assigned as the background, and the background subtracted signal at pixel $(x, y)$ in the $m^{th}$ image is $Im_{Bg\ sub}^m(x, y) = Im^m(x, y) - Im^1(x, y)$.

### [4.1.2] How to measure the signal?

Two possibilities allow measuring the optical signal from the pixels of the feature detected in the background substracted images. One can either consider the maximum intensity, brightest pixel (or minimum intensity for dark features), in the feature or the integrated intensity of the whole feature. For accurate optical intensity quantification, the integrated intensity is more correct. Indeed, objects smaller than the resolution appears as a diffraction-limited feature, of same projected size, whatever their real size. However, most optical microscopes are considered as linear instruments, meaning that a linear relationship exists between the cross-section of the object (*i.e.* its refractive index or mass, volume, $V_P$, or $V_P^2$ for scattering, amount of fluorophores, absorbance, etc…) and the observed optical intensity. When the cross-section is proportional to the number of molecules (fluorescence, Raman), the integrated intensity over the feature, $IntI$, is usually more accurate than the intensity of the brightest pixel, $I_{Max}$, for quantification.

Once the optical intensity of a feature or pixel is obtained it can be analyzed quantitatively depending on the type of measurement performed, and based on analytical expressions, some of which are proposed below. It is illusory to seek for absolute optical measurement. One would rather evaluate relative variations of a quantity (mass, volume, etc…) from optical intensity variations and resort to calibration procedures to verify the predicted trends, using different molecular probes or NP gauges, and cross-correlating the optical images with multiple microscopic observations [75–78].

### [4.1.3] Spatial resolution – the point spread function, PSF.

Resolution is the ability of an imaging tool to distinguish (resolve) in an image two objects. The resolution of a microscope is described by the Rayleigh criterion:

$$S_R = 0.61 \frac{\lambda}{NA} \qquad (42)$$

The shorter the wavelength and the larger the imaging lens $NA$, the better the resolution: for $\lambda = 532$nm (green light, approx. the average of the visible range) and 1.4 $NA$, $S_R = 230$nm.

It means that two point-objects in a sample closer than 230nm cannot be resolved. This means the edge of an object and any objects smaller than $S_R$ are merged in the image.

The image of the ultimate limit of a single light emitting point object, is given by a diffraction pattern with a central spot, named the Airy disk of radius $S_R$, which contains most of the emitted photons. The intensity profile of diffraction patterns is described by the point-spread function, PSF.

This reminds that a microscope never shows an actual object but only shows the projection of the object blurred (convoluted) by its PSF, usually described by a 2D Gaussian profile:

$$PSF(x, y) \approx I_{Bd} + I_{Max} e^{-\left[\frac{(x-x_0)^2}{2s_x^2} + \frac{(y-y_0)^2}{2s_y^2}\right]} \qquad (43)$$

centered on the point of coordinates $(x_0, y_0)$ and with $I_{Max}$ the maximum intensity at the centroid of the spot and, for symmetrical spots, $s_x = s_y = s_{diff} = \frac{0.21\lambda}{NA}$. The process of fitting intensity profiles in 1D, along lines, or in 2D, in $(x, y)$ planes, is named superlocalization. It affords different information illustrated in Figure 10.

### Conditions for fit

The image of a subdiffraction object should be represented by at least 3×3(×3) pixels (voxels in 3D). Superlocalization imposes to use objective with a minimal magnification. For a camera pixel of size $Cp$ and a resolution $S_R$, the lowest magnification that can be used for superlocalization is $M_{min} = \frac{3Cp}{2S_R} = 3Cp\frac{NA}{1.22\lambda}$.
From previous $S_R$ estimate, a CCD with $Cp = 6.5$µm requires ($M_{min} > 42$) a 60x oil objective.

### Superlocalization of objects' positions

If the validity of the Gaussian shape may be argued for fitting the PSF of asymetric entities (nanorods or strongly polar fluorophores, [79,80]), the Gaussian PSF provides accurate localization of the object centroid with 5-20nm resolution. This strategy is used to track with unprecedented resolution the motion (translational or rotational) of individual objects at surfaces during the course of a (electro)chemical process, such as micro or nano particles near or at electrodes by scattering-[78,81–83] or fluorescence [84,85] -based microscopies, or molecules confined in Raman hot spots [66], during their (electro)chemical conversion, or nanobubbles produced by gas evolving reactions [86,87]. More subtly, the superlocalization of the Airy disk highlights the centroid of the electric dipole of a nanoobject upon its electrochemical charging [83,88].
By tracking single objects diffusional (Brownian) trajectory in time and space, if necessary in 3D [89], the diffusion coefficient, henceforth hydrodynamic size, of the object can be inferred, helpful for NP growth/dissolution analysis [90,91].

### Sizing objects smaller than $S_R$

For sub-diffraction objects, the Gaussian PSF suggests the resolution is roughly the full width at half maximum, $FWHM$, of the cross-section of an optical feature: $FWHM = 2.355 s_{diff} = 0.49\frac{\lambda}{NA}$.
The objects size cannot be evaluated from the optical image, but their size (or the number of emitters) is at best inferred from the optical intensity, here $I_{Max}$, or from the integration of the optical spot, by comparison to predicted optical responses.

### Objects of size comparable to $S_R$

The edges of such objects may be seen as individual points, which gradually become distinguishable according to the Rayleigh criterion. This can be visualized by fitting the object PSF by a Gaussian whose $FWHM$ is larger than the diffraction limit, i.e. $FWHM > 0.49\frac{\lambda}{NA}$. A priori the object size can be estimated from $FWHM$, using for example models of optical images. In practice size variations are inferred from $FWHM$ variations.

### Objects comparable to larger than $2S_R$

Both edges of the object can be resolved individually and superlocalized, i.e. fitted by their own PSF. For objects presenting sharp edges, the inter-centroid distance represents the object size.

### Deconvoluting PSF in SPRM

In SPRM images (and also under some TIR conditions), the PSF of point sources is not disk-shaped, but presents a wave-like parabolic-shaped pattern (Figure 10) due to the direction of propagation of the surface plasmon. The parabola extents over >10µm along its axis and ~3µm in its orthogonal direction such that a diffraction-limited feature occupies a rather large area of an image. This poor resolution limits localization precision and imposes to image objects at low surface density. Single NP superlocalization requires post-treatment image deconvolution. The image of an object corresponds to the product of convolution of the PSF with the real spatial shape of the object that is revealed by the inverse operation using Fourier transforms (a convolution of functions is transformed into a simple product of the Fourier transformed functions). Such strategy, named deconvolution microscopy [92], was adapted to SPRM [93–96], to reach the localization resolution standards of most widefield microscopies and produce post-treated images with super-resolution rendering.

### [4.1.4] Axial resolution – the case of confocal microscopy.

Most often described in confocal microscopy, some widefield microscopy strategies provide information along the axial direction from a measurement of the optical phase of the imaged object based on interferometric principles developed in holography; they were used to track dynamically the electrochemical collision of single NPs [97]. Similarly, introducing phase delay in the axial direction distorts the PSF of an imaged object, enabling the estimate of its relative distance from the focus plane. This was used to monitor the approach curve of a SECM tip with 25nm axial-superlocalization resolution [98].

By definition CLSM has 3D imaging capability, and the pinhole used to block out-of-focus light improves the lateral resolution to $S_R = \frac{0.4\lambda}{NA}$ and axial resolution to $S_{Rz} = \frac{1.4\lambda}{NA^2}$ suggesting that a diffraction pattern is elongated by at least $2S_{Rz}$ =760nm along the axial direction. The drawback for high resolution CLSM imaging, besides high illumination power, is its slow imaging speed: a 150nm-resolution 512x512 pixel image (50nm pixel, 25.6x25.6µm$^2$) is acquired within 0.5s. Detailed protocols to optimize CLSM resolution imaging can be found in [99].

### [4.2] Transforming optical signal into electrochemical one.

If super-localization strategies allow probing through centroid position tracking $(x_0, y_0)$ dynamics of objects (position, size, shape, structure, polarization) associated to an electrochemical transformation, the local optical signal, $I(x,y)$ or $I_{Max}$, associated to the electrochemical transformation of a ROI, or of a single entity, on an electrode can also complement the electrochemical measurement. Taking advantage of the linear relationship between the optical signal, $I_{opt}$, and a cross-section $\sigma$, should provide a local estimate of the concentration of an optically active electrochemically converted species.

### [4.2.1] Electrochemical conversion of solution species.
**Direct optical probing of electrogenerated redox species**

The simplest situation corresponds to the probing of electro(fluoro)chromic species able to modulate light absorption (or light emission by fluorescence, or equivalently from Raman) following an electron transfer. Typically, in the $Ox + e \leftrightarrow Red$ reaction one of the species will be optically probed as it absorbs, emits or scatters a photon, the other being blind at the same wavelength.

When an electrode actuates the $Ox \leftrightarrow Red$ transformation, the spatio-temporal concentration profile of $Ox$, $C_{Ox}(x, y_{(,z)})$, or $Red$, $C_{Red}(x, y_{(,z)})$, can be inferred from the optical signal $I_{opt}(x, y_{(,z)})$ measured from optical images.

Two situations of interest are considered related to local light absorption or refractive index change. The unit voxel element of volume has the dimension of the image pixel in the $(x,y)$-plane, and of the microscope axial resolution, $S_{Rz} = \frac{2\lambda}{NA^2}$, along the $z$-axis.

The local refractive index of a solution, $n_{sol}$, is obtained, see (9), from the local concentration and molar refractivity of its different components, $Ox$ and $Red$,

$$n_{sol}(x, y_{(,z)}) = n_0 + \frac{1}{2}[C_{Ox}(x, y_{(,z)})R_{M,Ox} + C_{Red}(x, y_{(,z)})R_{M,Red}] \quad (44)$$

with $n_0$ the average refractive index of the solution in the absence of $Ox$ and $Red$. Formally, this equation includes all products consumed or formed, including the electrolyte and solvent contributions.

Assuming that $Ox$ and $Red$ have equal diffusion coefficient, the initial composition in the $Ox/Red$ couple, $C^*$, is preserved in the solution $C^* = C_{Ox}(x, y_{(,z)}) + C_{Red}(x, y_{(,z)})$, yielding a linear relationship between local concentration and local refractive index variations:

$$\Delta n_{sol}(x, y_{(,z)}) = \frac{1}{2}[R_{M,Red} - R_{M,Ox}]\Delta C_{Red}(x, y_{(,z)}) = \frac{1}{2}[R_{M,Ox} - R_{M,Red}]\Delta C_{Ox}(x, y_{(,z)}) \quad (45)$$

Useful in reflectivity-based optical microscopies, such equation was used to probe the electrochemical conversion of redox probes or HER (hydrogen evolution reaction) in the vicinity of microstructured electrodes or NPs by SPRM [100–102] and interferometric microscopy [103].

As SPRM probes ~200nm layer adjacent the electrode surface, much smaller than the diffusion layer, the local SPR signal reflects the electrode surface concentration:

$$I_{SPR}(x, y, t) \approx a[R_{M,Red} - R_{M,Ox}]\Delta C_{Red}(x, y, z = 0, t) + b \quad (46)$$

with $a$ and $b$ two constant parameters, allowing to transform local SPR intensity (local $n$ measurements) into opto-chronoamperograms or opto-voltammograms analogous to the electrochemical ones.

Otherwise, the optical signal is averaged over a large solution volume and one rather measures the average index of refraction over the diffusion layer:

$$n_{sol}(x, y, t) = \int n_{sol}(x, y, z, t) \, dz \quad (47)$$

Similar equations can be drawn for light absorption, considering the molar extinction coefficient of each species, $\epsilon_{M,Ox}(\lambda)$ and $\epsilon_{M,Red}(\lambda)$:

$$I_t(x, y) =$$
$$\approx I_0 \left(1 - 2.303 \int_0^d \left[C^*\epsilon_{M,Ox}(\lambda) + C_{Red}(x, y, z)\left(\epsilon_{M,Red}(\lambda) - \epsilon_{M,Ox}(\lambda)\right)\right] dz\right) \quad (48)$$

For luminescence emission (equivalently Raman) at wavelength $\lambda_{em}$, upon photon absorption at $\lambda_{exc}$, the local emission flux ensues:

$$I_{em}(x, y) \approx 2.303 I_0 \phi_L \int_0^d [C^*\epsilon_{M,Ox}(\lambda)$$
$$+ C_{Red}(x, y, z)\left(\epsilon_{M,Red}(\lambda_{exc}) - \epsilon_{M,Ox}(\lambda_{exc})\right)] \, dz \quad (49)$$

A similar approach extends at the liquid/liquid interface, for which similar derivations are proposed [104].

Such expressions allow transforming light absorption (luminescence or Raman emission) intensities into opto(fluoro)-voltammograms [105–107].

When the beam illumination and/or light emission is confined in sub-wavelength axial regions, as in TIRFM, this equation simplifies considering only molecules present within the evanescent field of thickness $\delta_{Ev}$:

$$I_{L,TIRFM}(x, y) \approx$$
$$2.303 I_0 \phi_L \left[C^*\epsilon_{M,Ox}(\lambda) + C_{Red}(x, y, 0)\left(\epsilon_{M,Red}(\lambda_{exc}) - \epsilon_{M,Ox}(\lambda_{exc})\right)\right] \delta_{Ev} \quad (50)$$

Confined optical imaging is also afforded by CLSM and a first approximation fluorescence collection yields:

$$I_{L,CLSM}(x, y) \approx$$

$$2.303 I_0 \phi_L \left[ C^* \epsilon_{M,Ox}(\lambda) + C_{Red}(x,y,0) \left( \epsilon_{M,Red}(\lambda_{exc}) - \epsilon_{M,Ox}(\lambda_{exc}) \right) \right] S_{Rz} \quad (51)$$

The number of photons collected by time unit is obtained by multiplying the former equations by the surface area of a pixel, $Cp^2$. Those expressions provide support to the linear correlation often observed between optical fluorescence, Raman [106,108,109] or absorbance-based [110] and electrochemical transients in studies combining an optical monitoring of electrochemical reactions.

To account for more complex optical configurations or geometries, a more rigorous analysis of the photon count is afforded by FEM. A full description is provided for 3D monitoring of an ECL reaction by confocal microscopy [111]. A complete modeling considers the contribution of the photons produced in the whole confocal cone together with those reflected from the electrode surface [109].

**Indirect optical probing – local pH sensing in electrochemical reactions**

Similarly the pH distribution over electrodes can be imaged operando using pH sensitive dyes. Those dyes are usually sensitive over 2-3 pH units centered around their pKa. The accessible pH range can be expanded by using mixtures of pH-sensitive dyes with different pKas [112].

The variation of the optical signal with pH is generally fitted by a sigmoid:

$$I_L(x,y) = \frac{B}{1+10^{a(pKa-pH)}} \quad (52)$$

where $B$ is related to the sensitivity (SNR) of the measurement, while $a$ reflects the dynamic range of the molecular sensor (larger pH ranges explored for $a < 1$). The law and its parameters are generally established from calibration curves, which have been validated under different microscopy configurations. More accurate pH profiles is also determined by FEMs [113].

The strategy applies to the indirect probing of a wide range of species involved in electrochemistry, such as small inorganic ions for which specific fluorescent molecular probes [114], or surface functionalized plasmonic NPs or SERS active molecular probes [115] have been described.

**[4.2.2] Probing the electrode-electrolyte interface.**

The same methodology applies to processes occurring for light active materials immobilized on surfaces. Previous expressions can be simplified in the absence of light-active material mass transfer. They are of the general form:

$$I_{opt}(x,y,t) \approx a C_{Red}(x,y,t)[A_{Red} - A_{Ox}] + b \quad (53)$$

with $a$ and $b$ constant terms and the $A_i$ terms corresponding to the molecular extinction, refractivity or scattering cross-section for a NP,…. They become for surface confined species at local surface concentration $\Gamma_{Red}(x,y,t)$:

$$I_{opt}(x,y,t) \approx a' \Gamma_{Red}(x,y,t)[A_{Red} - A_{Ox}] + b' \quad (54)$$

For a thin layer deposited on an electrode, $I_{opt}(x,y,t)$ is then related to the layer thickness, $d$, through its density, $\rho$, and molecular mass, $M$, as $\Gamma_{Red} = \frac{\rho d}{M}$.

This applies to reflectance-based microscopies for which a relative variation in reflected light, $\frac{\Delta I_r}{I_r}$, reports on relative reflectance variations, $\frac{\Delta R}{R}$. For thin layers, first-order expansion provides a local estimated of thickness (surface concentration)

$$\frac{\Delta I_r}{I_r}(x,y,t) = \frac{\Delta R}{R}(x,y,t) \approx a"d(x,y,t) \quad (55)$$

where a" is related to the refractive index of the solution, layer and substrate electrode.

The time-derivative of the optical signal is related to a local electrochemical current density inferred from optical images:

$$j_{opt}(x,y,t) \approx \frac{F}{A_{Red}-A_{Ox}} \frac{dI_{opt}(x,y,t)}{dt} \tag{56}$$

or for reflectance measurement

$$j_{opt}(x,y,t) \sim \frac{F}{a''} \frac{d\frac{\Delta I_r}{I_r}(x,y,t)}{dt} \tag{57}$$

Similar relationships can also be drawn for non-Faradaic processes, for example probed during electrochemical impedance spectroscopy. This is more amenable to optical microscopies sensitive to surface charge density, as those exploiting SPR [116,117] or refractive index variations. The general equations governing their quantitative description are given in [117]. They were applied in the quantitative imaging of single graphene [118], or Au nanorod [119] capacitance or to map the interfacial potential distribution at bipolar electrodes [120].

### [4.2.3] Quantitative measurements at the single entity level.

Granting access to quantitative physico-chemical measurement at the level of single entity is of major importance. Owing to the development of the electrochemical collision strategy [121–124], probing single NPs size, shape, electrocatalytic or chemical activity is becoming "routine". Moreover, the development of advanced electrochemical instrumentations and skills, such as low noise high bandwidth current amplification [124], sub-50nm nanoelectrodes [125,126], or nanopipettes *e.g.* for handling nanosized droplet electrochemical cells [127], or advanced fluidic nanogap electrochemical cells [128], has pushed electrochemical quantification limits close to the fundamental shot noise limit [129]. Though, the field lacks one of the great advantage of macroscale electrochemical studies: the possibility to challenge by *in situ* routine analytical characterization techniques the many mechanistic scenarios emanating from a 'simple' electrochemical curve.

Optical microscopy is an interesting helping hand if it is as quantitative as possible such that the optical signal readout correlates or complements the real electrochemical one. Two methodological situations are presented related to the information collected, either light scattering or luminescence, from individual objects.

**Quantitative information from localization and classification**

As was discussed above the localization of individual optical events provides a first methodological approach. It consists in recognizing and identifying events, classing them into behavioral categories and analyzing them from a statistical point of view. In this respect, machine learning methodologies (*e.g.* deep learning [130–132]) should soon help mining more massively the data acquired per (optical) image and accelerating their time-demanding processing, while removing some subjectivity bias.

Together with the spatial information on the localization of the event, kinetic information is also reached. Examples recently reviewed concern the fluorescence blinking of single molecules [133–135] e.g. for single (electro)catalytic turnover events at single NP, or the alteration of an optical signal (intensity, color, spectrum) during the (electro)chemical transformation of individual NPs or molecules at a nanocatalytic site [16,136,137].

**Quantifying single NP electrochemistry**

Some of the methodologies employed considering the scattering of light by NPs are detailed here. From the discussion in section A4B1C1, the optical signature of nanoscatterers can be rationalized by Mie theory for any type of NP and complemented by the Drude model for plasmonic NPs.

Aiming at a complementary *in situ* NP sizing from the intensity of scattered light, a variation with the square of the NP volume, $I_{sca} \sim V_P^2$, is expected; however, the electrode surface and the imaging conditions should alter this ideal trendline. Indeed various power laws, $I_{sca} \sim V_P^\xi$ with $\frac{1}{3} \leq \xi \leq 2$, depending on the microscope configuration (DFM, SPRM, IRM,…). Such power laws should be verified first on calibrated objects of the same composition confronted to same location SEM imaging and/or to optical modeling tools. An equivalent current or charge for the event of single NP growth or dissolution is given by[30,39,58,97,138]:

$$q_{opt}(x,y,t) \sim (I_{sca}(x,y,t))^{\frac{1}{\xi}} \quad \text{or} \quad i_{opt}(x,y,t) \sim \frac{d(I_{sca}(x,y,t))^{\frac{1}{\xi}}}{dt} \tag{58}$$

The (electro)chemical transformation of the NP affects principally its dielectric constant (and size). Optical intensity transients allow semi-quantitative NP composition, *e.g.* from optical modeling [31,139]. For plasmonic NPs, the conversion is related to changes in the LSPR characterized by the extinction peak intensity, $I_{ext}(\lambda_R)$, or wavelength, $\lambda_R$, evaluated by spectroscopy, which both are sensitive [60,139,140] to the NP composition (permittivity), the NP charge density, $N_e$, and the environment dielectric constant. As a rule of thumb, the LSPR condition according to Drude model predicts a resonance wavelength variation $\Delta\lambda_R$ with the surface charge $\Delta N_e$ or the surrounding permittivity, $\Delta\varepsilon_m$:

$$\frac{\Delta\lambda_R}{\lambda_R} = -\frac{\Delta N_e}{2N_e} \quad \text{or} \quad \frac{\Delta\lambda_R}{\lambda_R} = \frac{\Delta\varepsilon_m}{1+\varepsilon_m} \quad \text{or} \quad \frac{\Delta\lambda_R}{\lambda_R} = \frac{\Delta\varepsilon_m}{\varepsilon_\infty+\varepsilon_m} \tag{59}$$

Both relationships suggest the LSPR wavelength is a simple function of the electrochemical charge [140] or of the local electrolyte composition. The equivalent of an optical current is then estimated from its time derivative or similarly from its resonance peak intensity, a strategy used to draw single NP plasmon opto-voltammograms [29].

$$i_{LSPR} \sim \frac{d\Delta\lambda_R}{dt} \quad \text{or} \quad i_{LSPR} \sim \frac{dI_{ext}(\lambda_R)}{dt} \tag{60}$$

**Quantifying single molecule fluorescence from blinking events**

The blinking of molecular fluorescence is generally analyzed by single molecule localization fluorescence microscopy. It is depicted as successive cycles of on- or off-fluorescence emission at the same exact pixel localization. It then characterizes the activation or deactivation of fluorescence of an isolated electrofuorochrome generally adsorbed or confined on an electrocatalytic or electroactive site. It virtually represents a single electron transfer event.

The procedure is provided in Figure 11 for the case study of the reduction of the fluorescent methylene blue, MB, into its non-fluorescent leuco-MB form [141]. MB molecules are immobilized on a glass surface near a gold nanorod, NR. Owing to the NR LSPR and MB emission spectra overlap, the excitation of the Au NR LSPR by confocal laser triggers MB fluorescence. Meanwhile the reduction of a redox probe by a nearby working electrode regulates the redox composition of the surface-immobilized MB/leuco-MB. At high surface density, the average optical signal follows a Nernstian behavior with the electrode potential, $E_{el}$.

At the single molecule level, at a given pixel, successive transitions between on and off fluorescent states are observed. The distribution of the time periods in the on- or off-states ($t_{off}$ presented below) is fitted by exponential decay functions, revealing the average characteristic times, ‹$t_{on}$› and ‹$t_{off}$›, that describe the single molecule reaction probability (or kinetics). The ‹$t_{on}$›/‹$t_{off}$› ratio reflects the leuco-MB/MB ratio, also described by Nernst law.

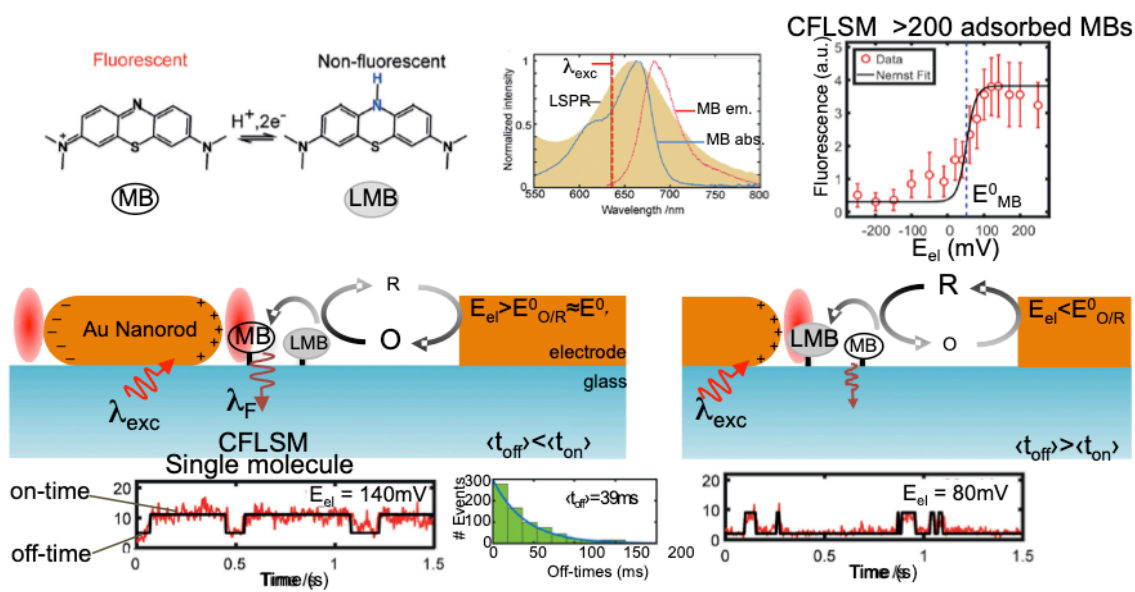

**Figure 11.** Confocal fluorescence microscopy of the redox switching at few to single adsorbed MB molecules. Fluorescence is enhanced by near-nanorod $\mathcal{E}$-field-enhancement. At the single MB level, $E^0_{MB}$ is obtained from $\langle t_{on}\rangle/\langle t_{off}\rangle$. Adapted with permission from [141].

This strategy is used in different configurations most often exploiting a confinement of the probing light either physically from the preferred adsorption of the fluorophore to a nanoparticle [133,142] or a nanobubble [86], or the confinement of single molecule into the restricted volume of a nanopipette [85] or an optical nanopore (E-ZMWs [71]) or by employing confined illuminations such as widefield TIRFM or confocal. These same strategies apply equivalently to SERS through molecular Raman vibration peak intensities.

**Correlated information**

A full correlation between optical and electrochemical signatures requires imaging the largest part of the field of the electrode. Different strategies have been employed. SPRM in the Kretschmann configuration provides lower resolution images but covers the widest 1mm$^2$ field of view. Any other imaging using high-NA microscope objective provides the highest spatial resolution but with a field of view at best 100x100µm$^2$. This is sufficient to map electrochemistry at classical disk ultramicroelectrodes [84,143,144] or nanoelectrodes [145–147] by fluorescence or label-free microscopies. These dimensions are accessible by microfabrication techniques [67,97,148] for producing microelectrodes that can be entirely monitored. An alternative consists in confining the electrochemical region of interest by a droplet of electrolyte held by a micropipette [149].

# [5] Selected Applications

Some examples where the optical microscope reveals features or mechanistic aspects that could not be seen from the single electrochemical measurement and provides a complementary, often quantitative, measurement of the electrochemical process.

## [5.1] Imaging the transport of species in solution

### [5.1.1] Down to the micrometer scale

The early works, inspired by Engstrom [3,150], proposed optical microscopies to probe and image the diffusion layer of electrodes to infer local or heterogeneous electroactivity, or evidence specific mass-transfer regimes. Diffusion layer of electrodes producing 'light active' (*e.g.* absorbing electrochrome or emitting electrofluorochrome) species can be probed quantitatively in 2D or in 3D at the diffraction limit by confocal fluorescence [151], ECL [111], Raman [108,109] absorption [110] or interferometric [103] microscopies, validating the theoretical concentration profiles under electrolysis conditions at very different electrode geometries, including SECM [152], or under forced-convection regimes [153]. This can be transposed toward electrochemical reactors, as fluorescence microscopy is a standard for monitoring macro to microfluidics. Fluorescent microparticle velocimetry, in flow channel electrolyzers, demonstrates how gas bubbles impact their performances [154]. Electrosynthesis within porous electrodes of quinone probed operando by fluorescence microscopy reveals dead end regions [155].

The diffusion of individual micron-sized 'light-active' objects towards ultramicroelectrodes, UMEs, can be treated similarly. The electrochemical current associated to the stochastic collisions on an UME of single fluorescent-labeled emulsion droplets can be correlated to a luminescence event (ECL) [156]. The sensitivity of the UME current hindrance to the landing position on the UME is revealed by observing the landing dynamics of insulating (fluorescent) beads [84,144].

Electrogenerated species are also probed with molecularly-specific fluorescent probes. pH-sensitive fluorescent probes (e.g. fluorescein) are largely used to map local pH (Figure 12A) encountered in various electrochemical configurations [113], including SECM [157], widened to more complex chemistry, *e.g.* bioelectrocatalytic processes [158]. Other fluorescent specific probes available from cell biology show potential interests: morin for $Al^{3+}$ in Al corrosion [159], calcium green for $Mn^{2+}$ release during $LiMn_2O_4$ battery operation [160]…

Bipolar electrochemistry provides an elegant alternative to detect the electrogeneration of non-fluorescent species [161,162]. The strategy consists in coupling a faradaic reaction of interest in one pole of a bipolar electrode to a fluorogenic reporting electrochemical reaction (ECL [163,164] or the electrochemical transformation of an electrofluorochromic species [165–168]) on the opposite pole. As long as the latter reporting reaction is not limiting, its optical imaging gives indirectly the rate of any "invisible" faradaic reaction of interest.

### [5.1.2] At the single NP level.

Observing the diffusion of single Brownian nanoparticles in solution towards an electrode (electrochemical nanoimpact strategy) can be achieved by light-scattering techniques and therefore by DF [82,89,97,169] fluorescence [85,170] or ECL-based [171] microscopies (Figure 12C). By tracking the NP trajectory, its hydrodynamic size can be inferred, complementing an electrochemical sizing. The strategy is then not only relevant for analyzing the NP transformation at the electrode but also to assess its conversion (growth [90,91] or dissolution [169]) in solution. Besides, the single NP trajectory analysis demonstrated that a faster NP phoretic motion could be triggered by different stimuli such as a local temperature gradient [172] or an electrochemical current flow [89,144].

### [5.2] Imaging the formation of products in electrocatalysis

In electrosynthetic processes, including (photo)electrocatalytic ones, the faradaic yield for the electrogeneration of a product is impeded by byproduct formation. Those are produced by kinetic competing routes usually inferred from macroscale current-potential curves, complemented by external (or *in situ*) titrations. However, the intricate micro or nanostructuration of electrodes allows tuning chemical selectivity [173], which supports the development of strategies able to monitor electrocatalytic processes at micro and nanoscales.

This section expands the previous one with the indirect imaging of electrocatalytic reaction products.

#### [5.2.1] At the micrometer scale

Bipolar electrochemistry can be extended to the monitoring of catalytic reactions. By using an array of bipolar UMEs, the heterogeneity in electrocatalytic reactions can be indirectly imaged by fluorescence microscopies (see Figure 12B). The images obtained should be quantitative as supported by the correlation between the current density flowing on one pole and the flux of photons emitted on the other pole [165,168]. This opens to electrocatalyst benchmarking strategies, so far with 6µm spatial resolution but which can be improved by using conductive nanofibers arrays.

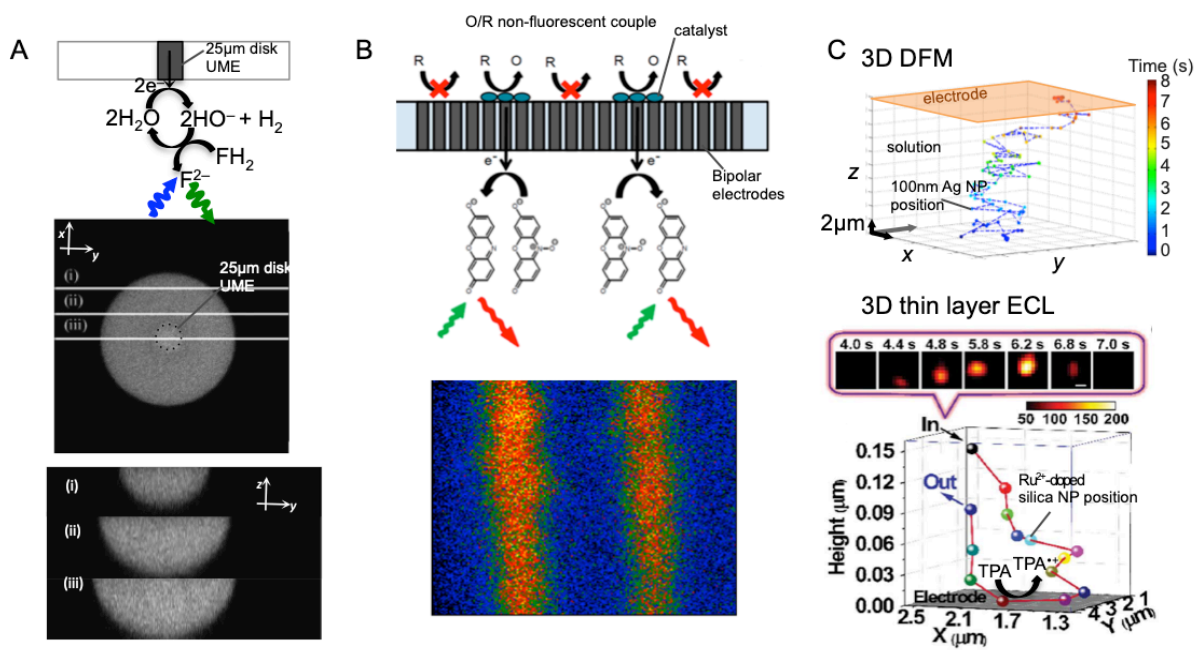

**Figure 12.** Imaging the transport of species in solution. (A) Direct 3D visualization by CFLSM of pH gradient electrogenerated during water reduction at UME (adapted with permission from [113]), or (B) indirect visualization of a catalytic "transparent" reaction using bipolar electrode array and an electroactive luminescent reporter (electrofluorochrome) in the lower phase; with permission from [165]. (C) 3D motion of single NPs to electrodes tracked (top) by holography from light-scattering (adapted with permission from [82]), or (bottom) without light illumination from thin-layer ECL; adapted with permission from [171].

Absorption and refractive index based microscopies are able to monitor electrosynthesis involving non-fluorescent reactants or products. The former requires probing strongly absorbing species within the thin diffusion layer adjacent to the electrode [110]. The second one evaluates the local electrode surface reflectance and is sensitive to molecular systems showing significantly differing molecular refractivities. Obviously the electrogeneration of gaseous molecules, with $n = 1$, can be probed (though most likely as gas bubbles, see below) and the electrocatalytic HER has been originally monitored by label-free SPRM at micrometric arrays of Pt NPs, down to visualization of the reaction at the single Pt NP level. It was the first quantitative analysis presenting an opto-voltammogram of HER [102]. Other molecules show refractive index differing from that of electrolytic solutions, *e.g.* formaldehyde that can be monitored in the operating conditions of methanol fuel cells. Its accumulation as a byproduct, quantified by SPRM, is due to the release of the electrode potential activating a self-catalyzed oxidation of methanol into formaldehyde by surface-adsorbed CO [174]. The contribution of adsorbed CO in the methanol oxidation mechanism was demonstrated by SERS [175].

However, SERS microscopy has a more recent development, mostly dedicated to the monitoring of the electrochemistry of model molecular systems presenting high Raman scattering cross-sections. Jain's group recently pushed SERS microscopy beyond model systems and devised catalytic systems relevant to $CO_2$ reduction or $O_2$ evolution in a biomimetic photosynthetic approach. Although these systems were explored under photocatalytic conditions, the strategy is sound for being transposed to electrochemical ones. The $CO_2$ photoreduction was studied in the gas phase at Ag NPs aggregates: a green laser excites the LSPR of Ag aggregates to induce $CO_2$ reduction. The ~$10^9$ electric field enhancement in the inter-NP hot spot provide single molecule Raman sensitivity. Figure 13 schematizes the imaging principle and the $CO_2$ reduction intermediate revealed spectroscopically dynamically. Of particular importance for future extension to electrochemical operando studies, stochastic $CO_2$ physisorption was probed dynamically owing to the unexpected $CO_2$ resonant Raman signature. Moreover, surface-adsorbed reduction products or intermediates, such as formic acid or CO, were also probed dynamically, revealing the phtotocatalytic activity of individual Ag NP aggregates [176].

Similarly, the green light photoactivation, in water, of photosystem PSII clusters adsorbed on Ag NPs aggregate enables the plasmon-induced photocatalysis of $O_2$ evolution reaction [177].

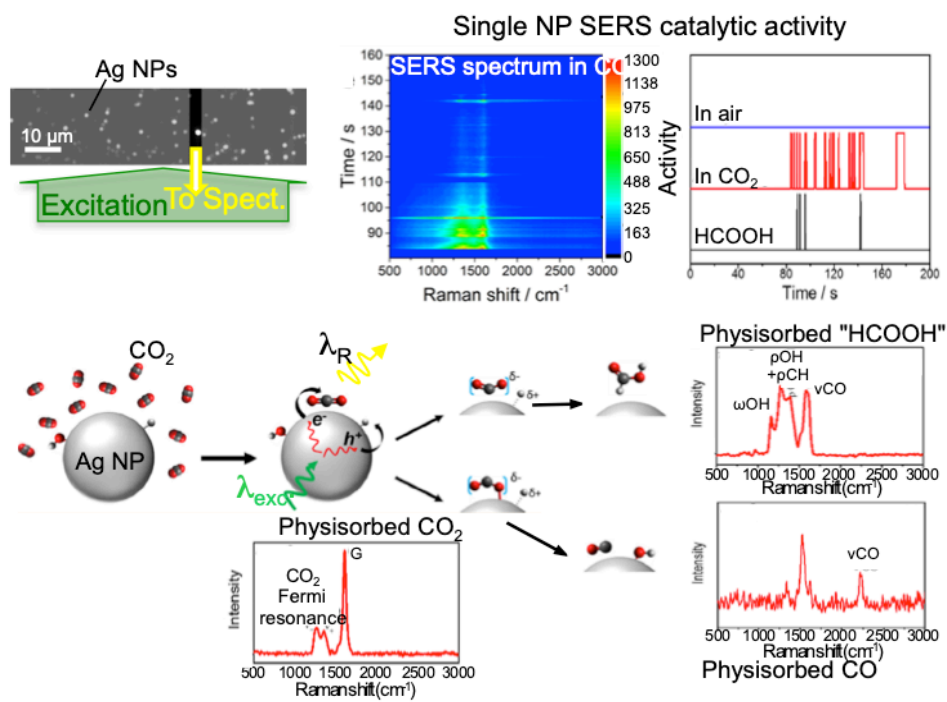

**Figure 13.** Principle of SERS imaging of $CO_2$ photocatalysis through $CO_2$ or reduced intermediates physisorption events at single Ag NP aggregates. Adapted with permission from [176].

### [5.2.2] At the single entity level.

Spectroscopic Raman identification complements other optical electrochemistry techniques to determine reaction potentials and rates at the single NP level. This can be reached by single molecule fluorescence microscopy and refractive-index based microscopies, either from variations of the scattered light intensity of the single NP during the electrocatalytic reaction, or, for plasmonic NPs, from variations in their scattering spectrum.

Singlemolecule localization fluorescence microscopy is indeed a powerful tool to investigate nanocatalysts in action at the single molecule and single NP levels. The technique, mostly used to study photocatalysis systems, with some examples related to electrocatalytic processes, was reviewed [178,179].

The fluorescence blinking strategy, presented in section A4B2C3, was used to address various single molecule oxidation or reduction processes. These studies take advantages of the rich redox chemistry of phenoxazine dyes such as the fluorescent resorufin, RS (Figure 14). Indeed RS is irreversibly produced, either from the reduction of resazurin, RZ, or the oxidation of amplex red, AR, while RS can be reversibly reduced to dihydroresorufin. This rich chemistry allows probing the catalytic activation of various oxidation or reduction processes at the single molecule level. The reduction of RZ by $H_2$ reveals the electrocatalytic activity of Pt nanoinclusions in graphene, and is a mean to apprehend their decaying performances [180]. The oxidation of AR by $H_2O_2$, a well-known strategy to probe the biocatalytic activity of various enzymes, is mature enough to dig into more diverse catalysts and reactions, such as the electrocatalytic ORR at $Fe_3O_4$ NPs. AR, by directly probing $H_2O_2$ formation at the NP, allows not only to quantify the 2-electron reduction route and also to infer the 4-electrons route contribution, and then to adress the activation or deactivation of the nanocatalyst [142].

Noteworthy, Nile blue belongs to the same family and presents strong Raman resonance and allowing to extend the strategy to single molecule tracking during its electrochemical conversion within Raman hot spots [181].

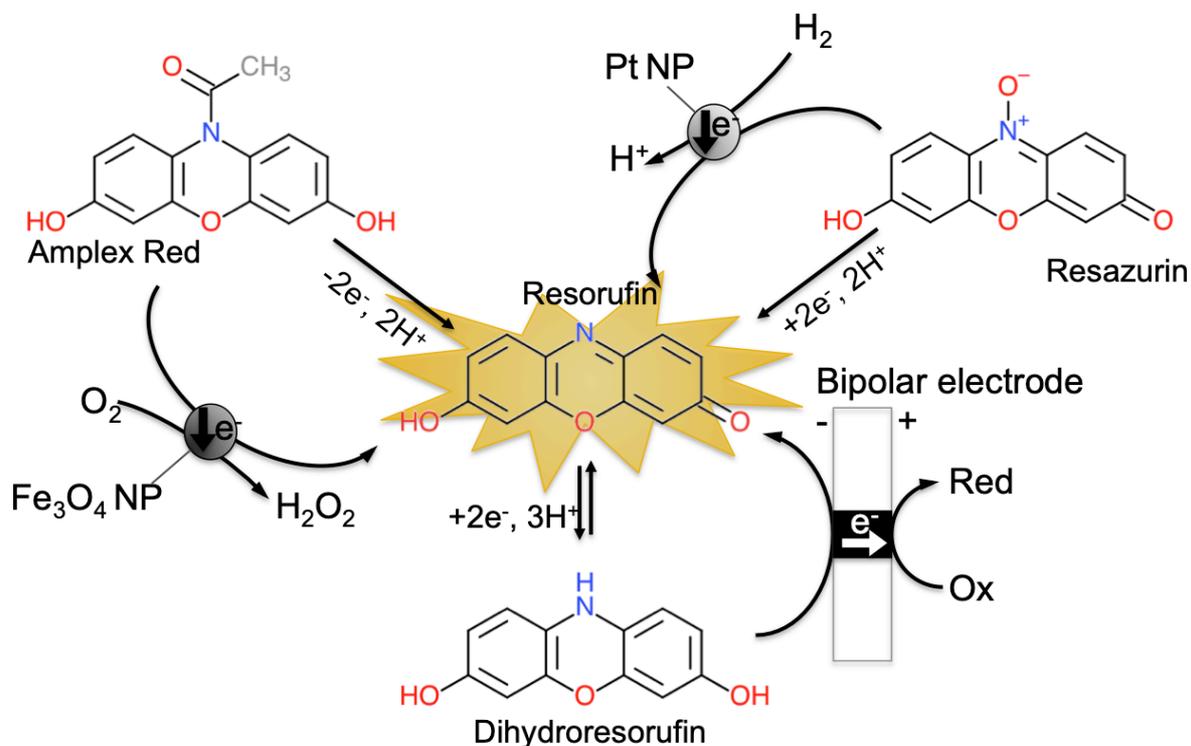

**Figure 14.** Redox chemistry of resorufin, from the phenoxazine family, allowing to image the catalysis of different oxidation or reduction processes using bipolar electrode (micrometer resolution) to single molecule fluorescence blinking at single NP.

The formation of reactants at the single NP level can be monitored optically at higher current densities. For example, for gas evolution reactions, large amount of gas with limited solubility readily saturates the electrode/electrolyte interface producing bubbles which can be optically revealed. The nucleation of a nanobubble, NB, on a 10nm nanoelectrode corresponds to a ~50 molecules detection, a challenge even for an electrochemical detection [126]. Bubble formation raises important issues in many electrochemical processes (corrosion, fuel cells, Al industry, small molecule activation,…), motivating recent interests.

Rhodamine 6G, at pM level, allows visualizing by TIRFM the adsorption of a single fluorophore at a single NB [86]. Each NB is then seen as a diffraction limited fluorescent spot (Figure 15) whose fluorescence intensity is an estimate of the NB size. The number of illuminated spots increases as the potential of the ITO electrode becomes more cathodic, in agreement with the increased HER rate. Taking into account the exponential decay of the evanescent field (the larger the NB, the smaller the spot intensity), the smallest NBs detected are 40 nm high. The strategy applies equivalently to the monitoring of NB formation on electrocatalytic surfaces such as gold nanoplates with higher NBs density generated at lower overpotential than on ITO.

Owing to its refractive index, $n = 1$, single gas NBs could also be revealed by refractive index-based microscopies. The dynamics of NBs growth and dissolution was evaluated by SPRM [182], evidencing by superlocalization the pinning of the NB on the surface. The formation of gas around plasmonic NPs can also be detected as a scattering intensity variation or a LSPR peak shift [183] suggesting that DFM is a meaningful tool for benchmarking electrocatalysts performances by counting the NB formation frequency [184]. However, this raises new issues: if the most efficient nanocatalyst allows growing a NB, the latter should rapidly disconnect electrically the catalyst from the electrolyte, halting gas evolution. Such effect was probed by IRM, that is able to visualize both individual Pt NPs and the growth of individual NB from individual NPs [185]. The optical modelling of the experimental configuration explains the peculiar variations of optical signals recorded from the images (Figure 15B). Such variations quantify the change in NB volume. Moreover the variation of FWHM of optical features probes the variation of the footprint of the NB on the surface, for footprints as small as 150nm. The NB footprint size and volume are deduced from the feature FWHM and intensity variations allowing a full 3D geometry (and contact angle) dynamica analysis of individual NBs. The smallest NB detected is indeed 30nm high, it is not hemispherical but spreads much more, with a 15-30° contact angle, onto hydrophilic ITO surfaces. The reason for this unexpected gas NB hydrophilicity, agreeing with recent single NB electrochemistry [126], is that a nucleus should accommodate high internal pressure (>36 atm inside a 40nm hemispherical NB). Increasing the NB curvature (lower contact angle) reduces by 5 folds the internal pressure. Being able to see both Pt NP and gas NB supports the coverage of the nanocatalyst during NB formation, stressing the need of mitigating it to optimize the efficiency of nanostructured electrode and to minimize overall energy loss.

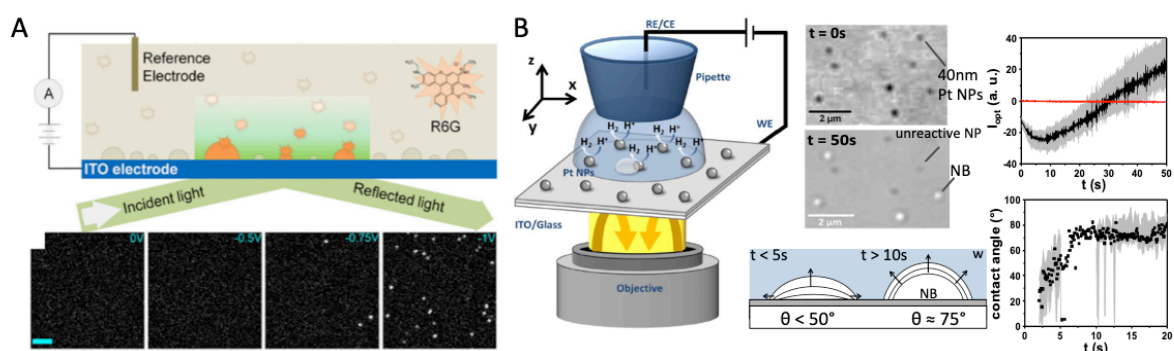

**Figure 15.** Imaging the formation of $H_2$ gas NBs (A) at ITO by single molecule TIRFM and (B) at single Pt NPs by IRM. (A) Principle and evolution of the number of NBs detected with the electrode potential; from [86], with permission. (B) The quantitative NB growth deduced from $I_{opt}$ variations allows estimating the dynamic evolution of the NB contact angle, $\Theta$, from [185].

### [5.3] Probing the electrochemistry of molecular adsorbates
Optical microscopies allow monitoring the influence of various molecular adsorbates on electrochemical processes. As in the previous section the different possibilities offered are illustrated going from the lowest µm resolution to the ultimate single entity observation.

**[5.3.1] Adsorbates and SAMs**

SPR is likely the most popular surface characterization technique able to detect molecular adsorption at unprecedented sensitivity, *e.g.* for immunassays. The potentiality of SPRM for imaging local electrochemical current density was first demonstrated at electrode heterogeneously coated with latent fingerprint (sebum) or thiol adsorbates [100] and submitted to cyclic voltammetry in a solution of a reversible redox probe while its surface was imaged. Local optical signals are evaluated in regions covered or not by the adsorbates (Figure 16) from which, local opto-voltammograms are reconstructed: in uncovered regions their current density is identical to the electrochemical one, while no faradaic current (down to 5pA/µm$^2$) is flowing in the coated regions.

Electroactive adsorbates are similarly imaged: spots of TNT adsorbates within fingerprint are imaged from the specific TNT electrochemical reduction, detected and quantified based on the local optical current measured.

For non electroactive adsorbates, local change in capacitance or in electrochemical impedance is more pertinent. Owing to the low invasiveness of electrochemical impedance spectroscopy and to the high sensitivity of SPR to charge density, SPRM imaging coupled to AC electrode potential was used to probe a large variety of non-electroactive (bio)chemical substances [186].

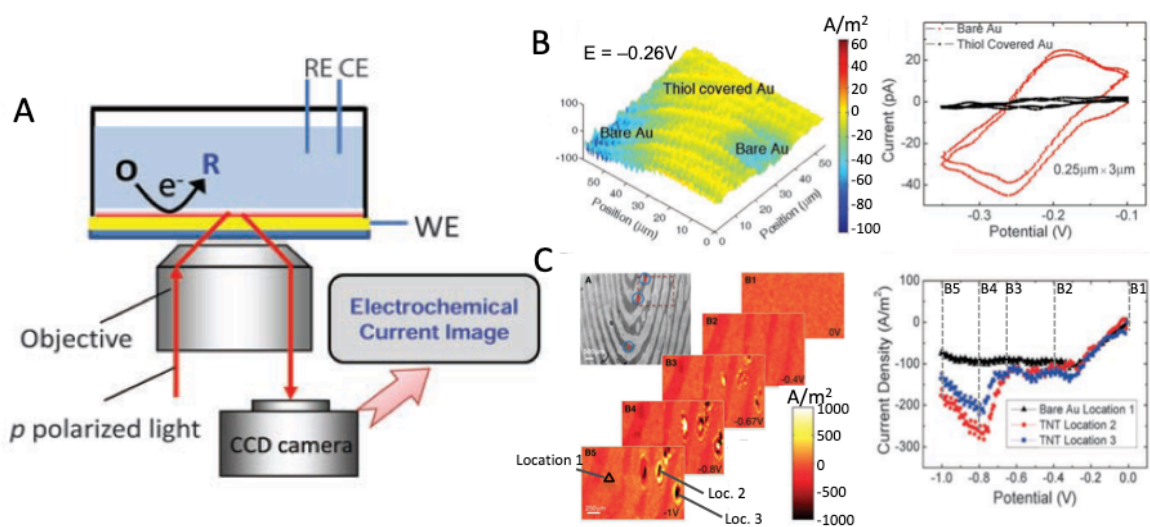

**Figure 16.** Quantitative SPRM imaging of the electrochemistry of molecular adsorbates on Au electrode. (A) Principle. (B) Imaging the heterogeneity of a thiol SAM from local SPRM-inferred voltammogram of a redox probe. (C) Imaging the electrochemical activity of fingerprint-coated electrode, showing local electron transfer blocking of fingerprint-coated regions and local reduction of TNT adsorbates (locations 2&3). From [100] with permission.

Because of to its lack of substrate generality, SPR is replaced in material sciences by ellipsometry to characterize solid surfaces and their adsorbates. Ellipsometric microscopy allowed monitoring at <0.5µm resolution the heterogeneity of non-electroactive self-assembled monolayers, SAMs, during electrochemical solicitation [187]. However it requires complex instrumentation and image post-treatment, with a sensitivity not better than a simpler normal incidence reflectance measurement, a strategy largely applied for the macroscale investigation of thin organic film on electrodes [188]. As a microscope, it allows imaging electrodes of many geometries and compositions. It was used to monitor *in situ* and real time the electrochemical growth of nm thick organic layers, obtained from diazonium salts reduction, on a 25µm disk UME. Local grafting rates were quantified, from Fresnel equations, with 0.2nm sensitivity and 0.5µm resolution over the microelectrode area. The higher rate towards the UME edges reveals the chemical instability of the grafted entity [143].

Fluorescence microscopy was also used to study the adsorption strength and dynamics of fluorescent molecular adsorbates on electrodes [189,190]. If many biological studies rely on fluorescence imaging of surface immobilized fluorophores, it is much more challenging when they are tethered to a metallic (gold,…) electrode, owing to strong fluorescence quenching by the metal layer. The shorter the fluorophore-electrode distance, the stronger the attenuation of its apparent quantum yield: it is <1% of the unbound value for <4nm. To observe luminescence from fluororescent-SAMs on Au, strong illumination is needed yielding its rapid turning off by photobleaching. Two strategies circumvent this problem: either the adsorbate is observed while it leaves the electrode (becoming more fluorescent), or fluorescence quenching lifetime is analyzed by CFLSM.

These studies exploit the reductive or oxidative desorption of SAMs [191]. Widefield fluorescence imaging probed the electrochemical desorption of fluorescent-SAMs covering a multi-single facets Au bead (Figure 17A). The onset potential for the SAM desorption depends strongly on local crystal orientation, providing an estimate of the adhesion strength of molecules in SAMs, correlated to the underlying atomic arrangement [192]. Besides evaluating the heterogeneity in SAMs surface density [193] and the nonspecific adsorption of molecular clusters, the strategy provides feedback for the preparation of functional SAMs by molecular replacement with the fewest possible defects [194]. Next, the response to CV of diluted SAMs of fluorophore-labelled DNA-tethers shows a potential-modulated elongation/collapse due to electrostatics (Figure 17B). Imaging this potential modulation on the multi-facets gold bead allows evaluating the influence of Au arrangement on both the DNA density and on the DNA reorientation switch characteristic frequency [195,196].

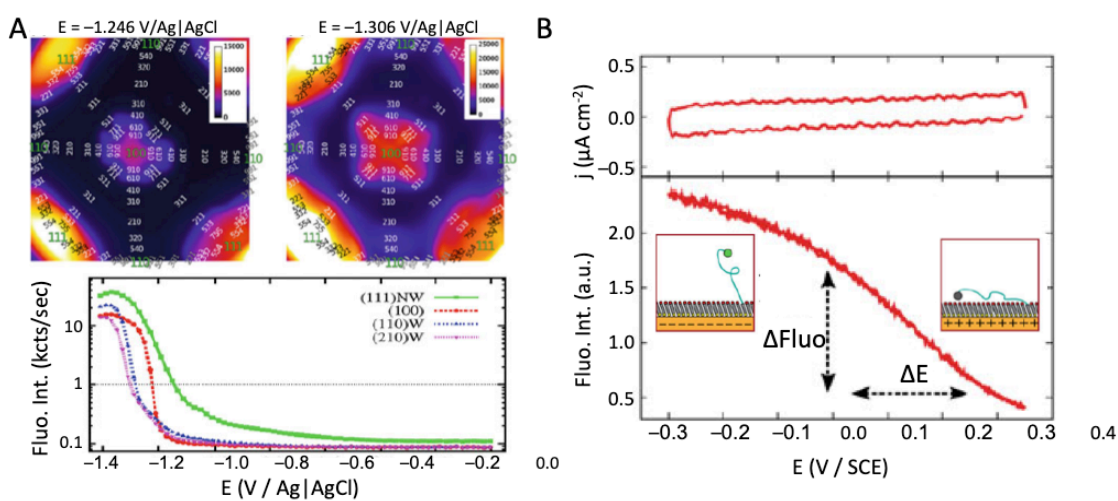

**Figure 17.** Fluorescence imaging of the electrochemistry of SAMs on Au electrodes. (A) Facet-dependence of SAMs electrodesorption illustrated from overlay of fluorescence images and electrode crystalline orientation map and fluoro-voltammograms. From [192] with permission. (B) Similarly the potential-induced orientation of diluted DNA in SAMs is probed from the fluoro-voltammogram, from [195] with permission.

ECL is an interesting alternative to photobleaching issues by prefering an electrochemical triggering of fluorescent events. The strategy was used in ECL-based biosensors for the sensing of various biomarkers immobilized on electrode or bead surfaces, some of them being brought to the market [197]. However, owing to the low efficiency of ECL, the imaging readout did not specifically seek for image resolution but rather for biomarker detection sensitivity. Since they only use electrochemistry as a mean to trigger the luminescence event, imaging-based bioassays are not discussed here; they are comprehensively reviewed [47,198].

The imaging of surface adsorbates by ECL microscopy was proposed in several studies where the fluorophore is either the adsorbate or present in solution.

Engstrom [2] imaged the heterogeneous electroactivity of an electrode surface from the ECL generation of fluorophore solutions. This strategy was extended to image sebaceous fingerprints by same-location dual-luminescence imaging: while ECL probes the electroactive regions of the electrode, photoluminescence complementarily reveals the sebum coated regions owing to luminescence quenching by the electrode [199].

Meanwhile, many ECL imaging-based bioassays preferably use polymers or beads functionalized or loaded with luminophores as sensing matrix, permitting to immobilize the luminescent tag near the triggering electrode. ECL images of these tagged arrays serve to sense and quantify biochemical relevant molecules: glucose or lactate, proteins, cancer biomarkers, cytochrome P450, or the reaction of metabolites with DNA [200]. Quantum dots in similar biosensor architecture opens the route to multicolor detection and therefore multiplexing [201].

One of the mechanistic explanation for the functioning of such sensors operating with labeled-beads relies on the chemically-evanescent light emission (Figure 7B). This chemical origin, demonstrated by SECM, was confirmed from the ECL illumination of only the first µm of a Ru-functionnalized bead adsorbed on an electrode oxidizing a TPA solution [50,51,171].

Both strategies using surface labeled or solution of fluorophores hold promise for new implementations to image the adhesion functions or permeability of cell membranes [202–204].

### [5.3.2] At the single entity level

Nanoscale resolution imaging was reached from the ECL imaging of the detection of 25nm conjugated polymer NP adsorbed on a polarized ITO electrode [205].

The ultimate level of single molecule resolution imaging is more amenable at surface-immobilized species. Since the impulse from [206] on the fluorescence imaging of the reduction and oxidation of single electrofluorochrome macromolecules, there is increasing interest in imaging the electrochemistry of single molecular adsorbates. This is reached mostly by single molecule localization by TIRF, SERS or TERS microscopies.

The electrochemical studies employing single molecule fluorescence microscopy were reviewed [207]. The general strategy is based on on/off fluorescence blinking of single molecule adsorbing/desorbing on single nanoobjects. To extend such study to non-fluorescing molecules, a competition is introduced in the catalytic scheme between a standard fluorophore and the molecule of interest. This is illustrated in Figure 18A for the photooxidation of hydroquinone, HQ (non fluorescent), evaluated through competition with AR photooxidation [208]. The photooxidation is performed by anodic polarization and the blue light illumination of a photocatalytic $BiVO_4$ truncated bipyramid particle. The cumulative traces of individual fluorescence events on the particle (Figure 18A) show stronger adsorption of AR at diagonal edges, probing also the electron transfer rates. Upon addition of HQ, its competitive photoxidation decreases the frequency of fluoresecence events, enabling local estimate of HQ photooxidation rates and revealing a higher adsorption and reactivity on the basal facet.

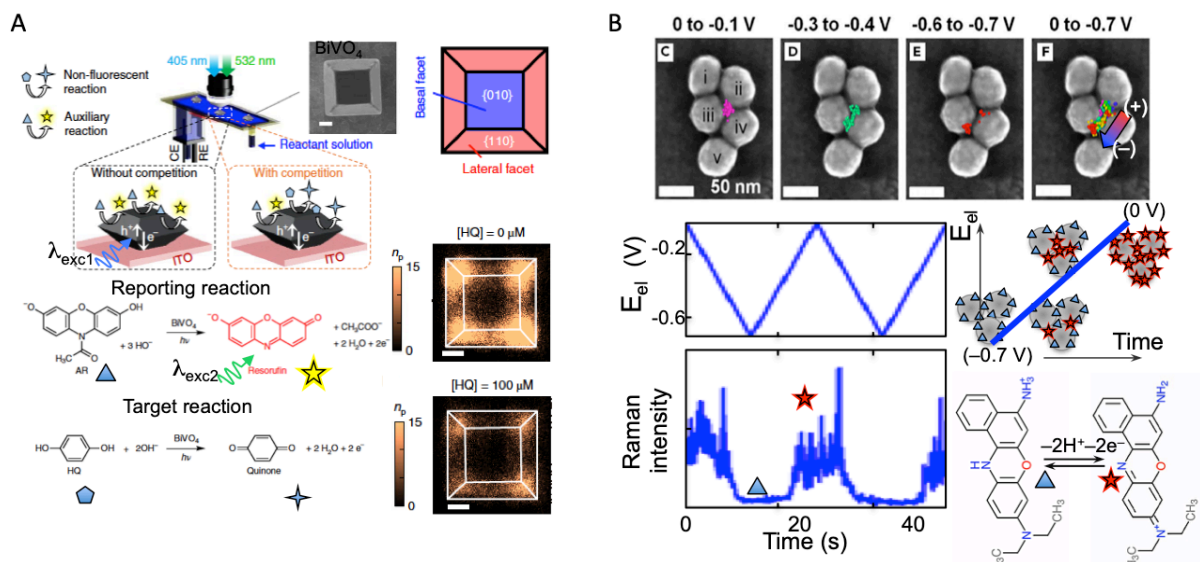

**Figure 18.** Super-localized adsorbates electrochemistry using (A) fluorescence or (B) SERS microscopies. (A) Strategy for evaluating the photooxidation of non-fluorescent probe (HQ) using a competition reaction. Adapted with permission from [208]. (B) Correlated SEM/SERS images of the potential dependent localization of Nile blue electron transfer on Ag NPs aggregates. Adapted with permission from [213].

SERS and TERS microscopies also afford single adsorbed molecule imaging together with a spectroscopic molecular identification. A tutorial explanation of the single molecule localization strategies employed in SERS is in [66], while the advances in the field of TERS and SERS are reviewed in [137].
SERS imaging of the electrochemistry of electroactive molecules at plasmonic NP aggregates have been discussed both at high (thousands molecules) and low (single molecule) surface coverage. Light is able to excite their LSPR, the photonic energy harvested produces hot carriers. These are transferred to the adsorbed molecules, and those present in hotspot regions presenting the largest electric field enhancement contribute to a detectable Raman signature. Most single molecule electrochemical SERS studies then deal with model electroactive molecules from which only one of the reduced or oxidized state absorbs light in the range of the plasmon resonance, such that a clear on/off Raman resonant signal is associated to the electrochemical transformation. The most studied molecules is Nile blue (a phenoxazine, i.e. resorufin analogue) or Rhodamine 6G. The electrochemical experiments consist in estimating the potential of these on/off Raman signal transitions during cyclic voltammetry.
An advantage over fluorescence microscopy is that if at each imaged pixel an integrated Raman scattering intensity is recorded, a complementary chemical signature is obtained through a Raman spectrum. Conversely, the slow acquisition (s range) of SERS spectrum is limited to slow kinetic analysis, compared to the ms range reached by fluorescence. This should be improved soon from recent report on 200Hz electrochemical-SERS probing by confocal Raman collection [209].

On the s range spectrum acquisition, SERS of single to few molecules provides insights on influence of the molecule-NP conformation on the charge transfer energetics. A SERS-inferred redox potential is typically assigned from the, gradual for large amount, or abrupt for single molecule, change in a characteristic Raman vibration intensity. The distribution of these potentials reflects both the Nernst potential and its variation related to the energetics of the molecule-NP interaction. Finally, the Raman inferred potential is correlated to the superlocalized centroid position within the NP aggregate highlighting the importance of nm range molecular motion in electron transfer processes [210–212]. SERS imaging is then a means to localize both the hot spot and the electrochemically active regions within NP aggregates (Figure 18B [212,213]), an information of importance for the development of plasmonic activation of (electro)chemical processes.

Unfortunately the performance of SERS is restricted to a limited class of substrates (nanostructured and plasmonic). Several strategies circumvent this limitation. One consists in covering a SERS active surface with a thin layer of the desired material, however, this "borrowing SERS" strategy [40] may not reproduce the nanoscale structure and composition of the most active catalysts. Similarly, shell-isolated plasmonic nanoparticles can act as Raman nanoreporters of structural changes within NP-electrode nanogap. SHINERS technique collects Raman spectra via a confocal microscope, but there has been no attempts of using it for Raman imaging of local electrochemical processes.

TERS microscopy affords an interesting solution to the lack of generality of substrate, also fueled by the promise of nm spatial resolution imaging, both in topography and in Raman signature, provided by a SPM nanotip. Since the works initiated by Ren [214] and Van Duyne [215], TERS microscopy has been more used to image electrochemical processes at surface-immobilized adsorbates.

In the model 2-electron, 1-proton reduction of ITO-adsorbed nitrobenzene, although the tip influences locally the ITO electrical double layer it does not perturb the electron transfer kinetics, setting grounds for EC-TERS. The measurement is performed within the limits of few to single molecules, allowing to elaborate upon the spatial distribution in the NB formal potentials for both the reductive and reversed oxidative steps. Particularly, the oxidized NB form shows a higher sensitivity to the local chemical environment. This expands to other electrode materials the importance of molecule-surface interactions in nanoelectrochemistry.

Ren [216] and others [217,218] implemented *in situ* visualization from the solution side, allowing nm spatial Raman imaging of electrochemical processes with <10nm resolution at wider range of electrode surface (*e.g.* opaque solids). If TERS imaging is still a delicate task owing to the difficulties in (i) reproducibly manufacturing the tip, (ii) avoiding its crashing on the sample during the long imaging process (>few h), (iii) avoiding its pollution by the spectroscopically probed adsorbate [219], these recent works have expanded the range of TERS applicability to more electrochemically relevant situations. They are now directed toward the understanding of the products formed during the irreversible electrochemical conversion of adsorbates: initiated with model systems such as hydroquinone [220] or nitrophenyl moieties oxidation [221], they address now electrocatalysis-relevant reactions such as the demetallation of iron phahalocyanine during ORR [222], or the plasmon-activated decarboxylation of carboxy-phenyl moieties [223]. Finally, the conversion of the electrode surface during an electrochemical process should be carefully considered: Au protruding nanodefects present on Au(111) terraces are preferentially converted into Au(I) oxide upon EC oxidation, while the terraces are mostly converted into into Au(III) oxides [224].

However, besides its inherently long imaging processing time, TERS is also often limited to molecules with rather large Raman cross-sections. A possible alternative can be reached with Scanning nearfield IR microscopy, successfully applied to characterize many materials. IR techniques inherently suffer from limited applicability in solutions, though an IR-compatible liquid cell architecture was proposed, analogous to *in situ* TEM liquid cell, with graphene sheets as impermeable walls [225]. It may find interest in near future.

### [5.4] Imaging the electrodeposition of conductive materials

The electrodeposition of metal or conductive materials, such as metal oxides or conducting polymers has been imaged in different ways from the microscale to the single entity level.

#### [5.4.1] Conducting polymers

The electrodeposition of conducting polymers onto an electrode gives it a color that can be optically imaged. The presence of sebaceous fingerprint on a metallic surface was probed by the specific electrodeposition of conducting polymer onto the sebum-free regions of the electrode. The electrochromic properties of conducting polymer are further used for image contrast enhancement [226].

Owing to strong local change in refractive index, the electrodeposition of polyaniline at single plasmonic Au NPs can also be imaged by hyperspectral DFM complemented by Raman microscopy. The redshift in the scattering spectrum of the LSPR of the Au NP provides an indirect estimate of the deposited polyaniline thickness (Figure 19) that is correlated to the aniline oxidation current. Illuminating the NPs with laser ligth for different wavelengths demonstrates the plasmon-mediated activation of the electrochemical reaction. Indeed, only the 561nm excitation of the LSPR photogenerates hot holes facilitating the aniline oxidation by 0.24V (10% of the illumination light energy). This study [227] shows the promise of optical microscopy for evaluating plasmon-mediated (electro)catalytic processes, while this emerging field of electrochemistry was so far only addressed by ensemble measurements [228].

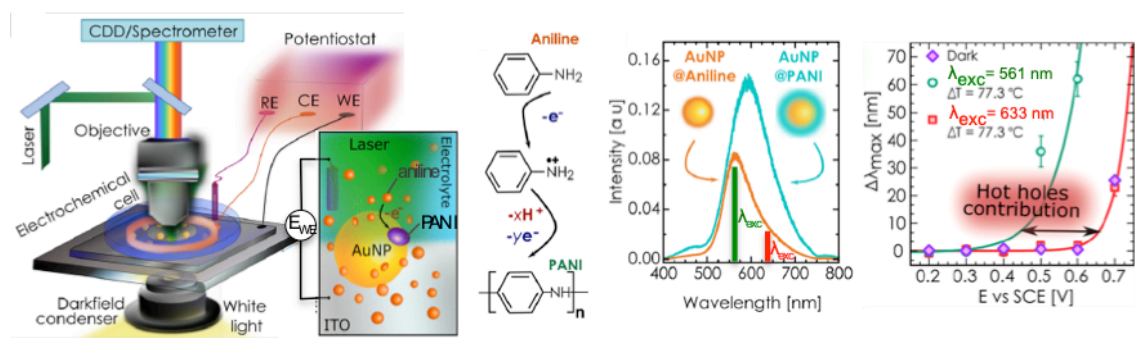

**Figure 19.** Plasmon-mediated polyaniline, PANI, electrodeposition on single Au NP triggered by laser LSPR excitation and probed by hyperspectral DFM. Adapted with permission from [227].

**[5.4.2] Metal deposition**

Many in situ techniques, among which optical ones, have provided valuable complementary information to electrochemical methods in the understanding of metal electrodeposition, from the early nucleation stage to the metal phase growth. Going from lowest to highest resolution, the long time electrodeposition monitoring is invaluable in the understanding of most rechargeable metal anodes. This issue is apprehending and remediating the formation of dendrites upon charging a battery. Bright field optical microscopy, with sub-mm resolution, was used to monitor operando in miniaturized battery analogues, the dendritification of Li [229] or Zn [230] anodes during charging. The origin of dendrite formation is, unexpectedly, the onset of the electrolyte diffusion limitation within the mossy electrodeposited Li.

The understanding of electrodeposition processes may also come from a microscale visualization of its earliest stages. At high-energy surface electrodes, such as noble metals, the electrodeposition of metals starts by an atomic layer deposition, followed by three-dimensional growth. *In situ* reflectivity or ellipsometry provides a real time estimate of the film thickness [231], based on the formalism recalled earlier. At low energy surface electrodes, such as carbon or ITO, the electrodeposition of metal is initiated by the stochastic appearance of nuclei which results in particles of broad size distribution. A comprehensive review of the nucleation-growth models applied in particle electrodeposition is found in [232]. Monitoring electrodeposition *in situ* or in real time allows (i) controlling the growth size and direction of electrodeposited particles [233], and (ii) challenging the proposed models by atomic resolution electron microscopies [234–236]. Optical microscopies, probing the scattering properties of metallic NPs, offer a simple operando wide field (50Hz, 50x50μm$^2$) methodology for electrodeposition imaging at high throughput. This was case proofed for the electrodeposition of Ag NPs, monitored by DFM [30], or IRM [58] whose images carry quantitative information on the electrodeposited NP size, with detection limit down to 45 or 10nm, respectively.

The IRM monitoring of the cathodic electrodeposition and anodic redissolution of Ag NPs during a cyclic voltammetry, CV, experiment provides opto-voltammograms showing the growth and further dissolution at the single NP level (Figure 20A) approaching the ensemble electrochemical response. The broad NP size distribution generated during the cathodic electrodeposition step, together with the high sensitivity of the optical monitoring, provides further ground for quantifying from a same experiment the intervention of the deposited metal surface tension on the NP electrodissolution potential (the smaller the NP, the easier the oxidation).

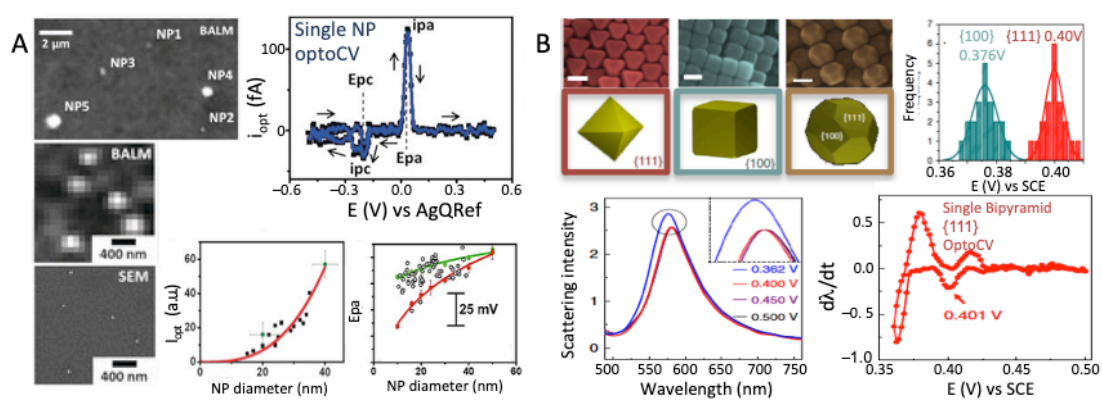

**Figure 20.** Single NP optical voltammetry for the electrodeposition of (A) Ag NPs on electrode and (B) Ag atomic layers on 100nm Au NPs. (A) IRM images of different size NPs

allowing, with same location SEM, correlated $I_{opt}$-size relationship; ensuing quantitative CV reconstruction for the growth/dissolution of a 20nm NP and size-dependent electrodissolution, adapted from [58] with permission. (B) Separating the facet-dependent UPD energetics from spectroscopic DFM. From [59] with permission.

The versatility of optical microscopies enables probing electrodeposition processes at varieties of supports. DFM illumination of the apex of a nanoelectrode allows detecting the electroeposition of single particle from the variations in the light scattering at the apex. This was demonstrated for the cathodic deposition of Co or Au NPs. From the analysis of the scattered light (i) NPs are *in situ* dynamically detected down to 65nm radius for the non-plasmonic Co metal, and (ii) mechanistic insights is obtained from correlated EC and opto-voltammograms. Particularly, the reduction of $Co^{2+}$ ions shifts from Co(0) deposition to its conversion into Co oxide owing to a catalytic involvement of Co(0) in the ORR.

The ultimate imaging resolution is electrodeposition at a single NP. LSPR is highly sensitive to the deposition of an external shell of an extraneous metal (Ag [59,237], Hg [238,239]) on a variety of geometries of Ag or Au NPs, which were imaged at the single NP level by spectroelectrochemical DFM. The metal deposition is detected as up to 100nm blueshift in the NP LSPR wavelength.

In its most sensitive configuration, the electrodeposition of sub-monolayer of Ag onto single Au nanocrystals was demonstrated. Opto-voltammograms describing the electrodeposition are evaluated from the shift in the Au NP LSPR (Figure 20B). Beyond the common overpotential electrodeposition, the less cathodic underpotential, *i.e.* single atomic layer, deposition, UPD, appears, through ~1-2nm LSPR blue shift, with a sensitivity of few thousands of Ag atoms. The study of different NC geometries further demonstrates the easiest UPD on the (111) facets than on the (100) ones.

Noteworthy, the deposition of metal shell onto plasmonic NP can also be operated at the single NP level in the absence of electrochemical source, *e.g.* via galvanic exchange reactions under various conditions. Though not purely electrochemical systems, similarities with electrodeposition is found: the ensemble NPs slow transformation kinetics does not reflect the fast single NP one, but rather the large distribution of the stochastic induction time [240].

### [5.5] Imaging the electrochemical conversion of solids

The conversion of electrode or solids is first described between chemically stable states. Often the electrochemical conversion of materials consists of the formation of oxides or their dissolution in an electrochemical environement and rather relates the impact of corrosion on electrochemical performances. The conversion of materials in the presence of halide anions, reputed corrosion activators, is then discussed.

#### [5.5.1] Probing reversible conversion processes
**At the micrometer scale**

Most electrochemical conversions of the active material used as electrode for battery applications are associated to change in color (electrochromism), which can be probed by absorbance measurements in transmittance brightfield microscopy.

2D materials are attracting considerable interest in electrochemistry. Graphene or reduced graphene oxide, rGO, or $MoS_2$, have been proposed as promising electrode material in different energy storage applications. Upon reductive intercalation of Li or Na, rGO [241] or MoS$_2$ [242] become more transparent, enabling a simple operando evaluation of their conversion from transmittance BFM, mostly at mm to 10µm resolution, with complementary micro-Raman or photothermal imaging [243].

High voltage electrochemical exfoliation is a means to produce 2D graphene layers but with quality strongly dependent on their degree of oxidation (graphene oxide, GO, is poorly conducting). Monitoring the graphene oxidation process is pertinent and possible by various optical microscopies, owing to its exceptional optical properties. A monolayer graphene sheet deposited onto a glass slide is submitted to oxidation/reduction cycles while images by IRM. Upon >1.4V oxidation, micrometer-sized flower-like patterns (Figure 21), propagating over the graphene surface, mostly from defects, are assigned, from same location micro-Raman, to local transformation into GO. IRM provides an operando quantitative imaging of the graphene degree of oxidation. It enables confronting the chemical (cGO) and electrochemical (eGO) production of GO. If the electrochemical conversion is produced by much larger oxidation rates it is also reversible, as eGO is reverted to graphene (from IRM images) by electrochemical reduction. The role of reactive oxygen species is also diagnosed further allowing to evaluate the 1e-1H$^+$ stoichiometry of the graphene oxidation [244].

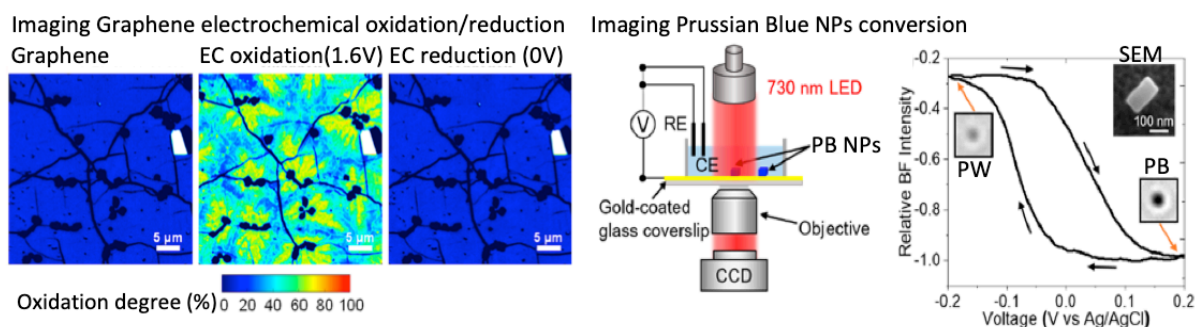

**Figure 21.** Left: IRM visualization of the reversible oxidation↔reduction of Graphene↔GO; adapted from [244] with permission. Right: Bright field transmittance imaging of single PB↔PW nanocube electrochemical conversion. From [246] with permission.

### At the single nanoparticle level

The electrochemical conversion of absorbing or fluorescent molecular dyes encapsulated in sub-pL emulsion or droplets confronted to individual opto-voltammogram allows scrutinizing the mass-transfer/partitioning driven electrolysis within such nanoconfined environment [170,245]. Hexacyanoferrate crystals such as Prussian blue, PB, are another electrochromic material with applications in energy storage or sensing. Upon reduction, accompanied by intercalation of one K$^+$ per electron, a red light absorbing PB nanocube is transformed into transparent Prussian White, PW, as probed by BFM (Figure 21) and SPRM. The single PB nanocube opto-voltammogram suggest a rather slow conversion owing to K$^+$-transport limitation. It is also highly stochastic, as some nanocrystals are not reactive or show low cyclability. Sputtering a thin layer of Pt atoms over the electrode assembly electrically connects all nanocubes that now present excellent cyclability [246]. It highlights that a majour source of stochasticity in single NP electrochemical response should be released by a careful control of the electrical contact as well as the NP-electrode interface (surfactant, oxide layers,…), a topic that should be emphasized in this field.

Metal oxides also often show electrochromism, such as $WO_3$ [247], Fe or Co oxides. Moreover, the latter abundant transition metal oxides are interesting alternative to rare earth metals electrocatalysts, used as positive electrodes in batteries or anodes in electrolyzers. The electrochemical behavior of Co oxide, CoOx, was probed by both SPRM and DFM. The decrease in electron density upon oxidation results in a decrease in the CoOx refractive index, and CoOx oxidation is probed as a decrease in scattering (then SPR) intensity. In a battery-related context, the opto-voltammogram inferred from SPR during the oxidation of single $LiCoO_2$ NP shows a reversible delithiation/lithiation correlated to the reversible ensemble EC response. As the SPR signal probes preferentially the $LiCoO_2$ component, it provides an estimate of the delithiation rate, complementing the electron transfer rate obtained from the EC response [248,249].

Since CoOx are self-healing electrocatalytic materials, a single CoOx sub-microparticle was electrodeposited on a nanoelectrode to collate both single EC and DFM response associated to CoOx particle oxidation [146]. The opto-voltammogram obtained from the scattered light variations reveals the Co speciation (Co(II) vs Co(III)). Compared to the EC response, it evidences the contribution of catalytic water splitting by Co(III)Ox. Meanwhile, the superlocalization of ~1µm particle edges with resolution down to 20nm demonstrates that the CoOx oxidation/reduction is accompanied by a large (>30% in volume) and reversible particle breathing (expansion during Co(II)Ox oxidation).

### [5.5.2] Probing corrosion processes
**At the micrometer scale**

Macroscopic corrosion of metals results from subtle chemical environments that can be only probed at the microscopic level. These localized corrosion processes initiate on discrete locations through galvanic couplings; this is an ideal playground for *in situ* techniques, including optical microscopies.

Generally, the local anodic dissolution of a metal is associated to a local pH decrease, while pH increases in a nearby cathodic region upon local oxygen reduction. Local solution alkalinization can be probed by fluorescence microscopy with fluorescein. CFLSM imaging of an Al 2024 surface reveals the cathodic behavior of the regions surrounding Cu-rich intermetallic inclusions. The fluorescent molecule is actually not detected in solution but entrapped into an Al oxyhydroxide precipitate deposited on the corroded Al matrix surrounding the cathodic inclusions. These lesser-reflecting and protruding deposits are complementarily imaged by reflectance-CLSM [250] or at higher resolution with near-field optical probe, NSOM [251]. The latter, by tip-surface distance control, allows imaging at high-resolution the topographic changes associated to local Al dissolution and redeposit of Al oxyhydroxyde. These *in situ* revealing modes of corrosion are expanding either as optical sensors [159,252] or incorporated in "smart coating" [253].

Diffraction-limited microscopies enable imaging corrosion at its earliest stage. The localization of pH-dependent fluorescence events allows counting the pitting sites in iron corrosion, with further confocal Raman identification of corrosion products [254]. Reflectivity microscopy was used to image the formation of iron oxide layer during iron corrosion. The process is again highly localized, allowing to identify the most active regions of the surface from local opto-voltammograms revealing micrometric regions of iron oxide formation/dissolution, an information buried in the multiple contributions averaged in a polarization curve [20].

**At the single NP level**

Besides NSOM for surface topography images, pure optical topography is afforded by differential or dual-beam interferometric microscopies [103]. Implemented in a CLSM, differential interferometric detection can image ultraflat Au(111) monocristal surface with monoatomic steps resolution, enabling a dynamic imaging of its localized anodic electrodissolution [255].

The behavior of metallic NPs to strongly oxidizing conditions provides insight into their corrosion resistance. The oxidation of metal NPs results either in their dissolution or the production of a thin oxide layer shell. In the presence of precipitating pseudohalides ($Cl^-$, $S^{2-}$, $SCN^-$,…), metal NPs oxidation results in accelerated/slowed dissolution or their transformation into ionic crystals (core-shell NP at the earliest stage). The insights gathered by optical microscopies in the most studied case of Ag NPs are summarized in Figure 22.

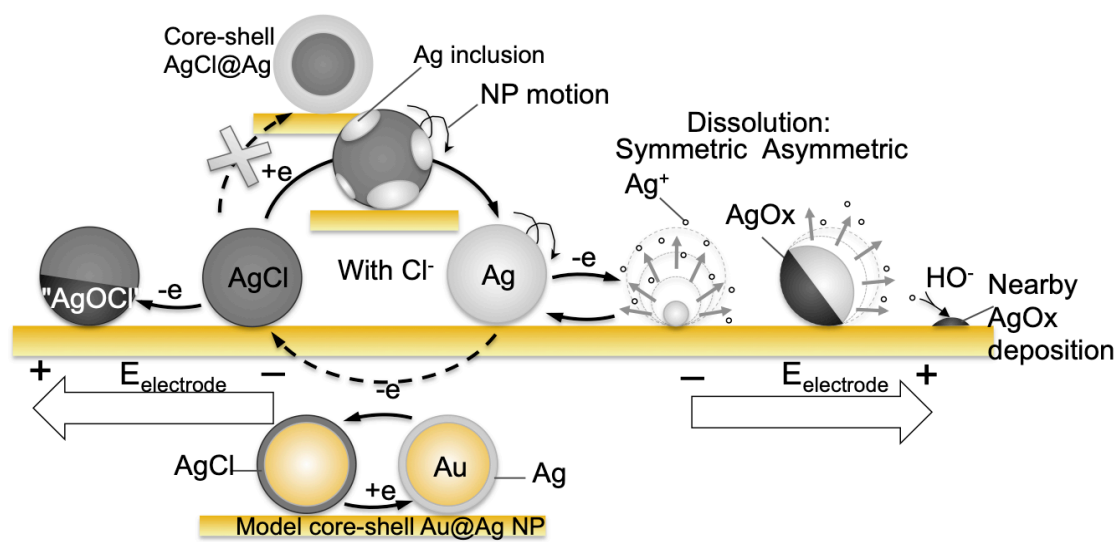

**Figure 22.** Examples of NP intermediates and mechanistic insights probed by optical microscopies during nanosilver electrochemistry.

The oxidative electrodissolution of metallic NPs is the simplest to study optically, from a decreasing optical (scattering) signal. This was detailed in number of configurations for Cu [138] or Ag NPs, either electrodeposited [30,58], or colloidal NPs pre-adsorbed [60,61,138] or freely diffusing [39,82,85,97], in solution during their stochastic collisions with a polarized electrode. The dissolution kinetics is quantitatively assigned at the single NP level based on the correlation between the optical signal variation and the EC one. If the oxidation of immobilized NPs results in a complete dissolution, *e.g.* following a bell-shaped opto-voltammogram, the mobility of NPs near the electrode surface rather results in stepwise oxidation. This motion, tracked in 2D or 3D, by fluorescence, DF or reflectivity microscopies evidences NP bouncing, adsorption, or surface diffusion on the electrode. The 'hit and run' scenario, suggesting multiple oxidation steps for a same NP, confirmed by high frequency electrochemical experiments [256], is an important path to consider in further single NP electrochemistry.

In the presence of Cl$^-$, the CV of individual surface-confined Ag NPs follows a reversible oxidation into AgCl NPs, evidenced by the changes in their LSPR spectrum [60,257]. Indeed, the formation or presence of a shell of dielectric material surrounding a plasmonic core induces a ~10nm LSPR redshift (smaller than for a metallic shell). Insights into the Ag↔AgCl conversion are provided at the nanoscale using a model Au core Ag shell, Au@Ag, NP [139]. From LSPR, the EC conversion of the Ag into AgCl shells is reversible, and proceeds by propagation of the Ag/AgCl interface bteween the Au core and the electrolyte interface. It is confirmed from the optical identification of reaction intermediates, at the level of NP-dimers, and supported by optical models: LSPR spectrum reveals a transition between a dimer of capacitively coupled Au@AgCl NPs to a strongly resonant mode of conductively coupled NPs in a Au@Ag dimer.

The slow conversion of AgCl raises the question of the full conversion of AgCl NPs. This is addressed from the IRM investigation of the reduction of individual surface-immobilized colloidal AgCl NPs [31]. Optical models suggest that IRM can be quantitatively used to differentiate the AgCl reactant and Ag product NPs, as well as the presumed intermediate reduction products. Owing to its poor electronic conductivity, AgCl NP reduction proceeds by the slow and stepwise formation of multiple conducting inclusions, facilitated by the NP rolling over the electrode surface, rather than by a core-shell mechanism.

At more anodic potentials, towards the water splitting region, the oxidation of AgCl results in a further redshift and bandwidth broadening of LSPR, evidencing the formation of a stable AgOx NP [60]. In the absence of Cl$^-$, the same AgOx formation competes with the Ag NP dissolution. The formation of AgOx is complementarily evidenced from the attenuation of the Ag NP fluorescence at high anodic potentials [258]. Further insights into the competition between Ag dissolution and AgOx formation are obtained by super-localization DFM [88]. The process is associated to a ~40nm displacement of the optical feature centroid revealing the asymmetric dissolution of the NP and formation of poorly conducting AgOx patches. At even higher potential, the combination of OER and Ag dissolution produces AgOx nanoclusters whose accumulation nearby the original Ag NP seed is imaged [259].

The knowledge and control over oxide layer formation at noble metal is of importance for apprehending their electrocatalytic activity. The formation of AuOx shell is detected by LSPR redshift [260] and concomitant damping of the LSPR intensity [261]. Its influence on the catalytic activity of Au NPs, for either $H_2O_2$ or $N_2H_4$ oxidation, was evaluated from LSPR shift upon addition of these coreactants. The $N_2H_4$ oxidation is dominated by surface reactions: AuOx formation, ligand removal or potentially nanobubble formation [149]. Interestingly, $H_2O_2$ oxidation removes some of the AuOx formed. Cl$^-$ ions play a role in catalysis and corrosion at Au, and their intervention as surface adsorbates on AuOx blocks $H_2O_2$ oxidation [260].

### [5.6] Probing non faradaic processes

The contribution of the electrochemical double layer, EDL, can also be probed by optical microscopies. The change of double layer structure with a metal electrode potential induces a change in its macroscale optical reflectivity (electroreflectance) [231,262,263]. The imaging of the influence of ion adsorption or ion cloud distribution in EDL charging is a challenging question as it involves small variation in surface charge and thus requires sensitive detection modes. At plasmonic NPs, ion adsorption results in >10nm shift and broadening of the LSPR peak while EDL is more subtle (~few nm LSPR shift according to the Drude model). If modest for plasmonic nanospheres, stronger contributions are expected in future studies at supercapacitor nanomaterials. A SPR optical fiber sensor was indeed proposed to quantify the charging capacity of a supercapacitor $MnO_2$ nanosheet [264].

By order of decreased sensitivity, the process of ion adsorption is first discussed. Ion adsorption has a major contribution on the overall electrocatalytic performance of nanocatalysts which can be probed at the single NP level by DFM spectroelectrochemistry. Opto-voltammograms, named single particle plasmon voltammograms, spPV, established from the LSPR peak intensity, identify the importance of ion adsorption on the electrochemistry of Au NPs or NP-dimers, for higher sensitivity. As ions adsorb on the NPs the LSPR scattering peak is redshifted and broadened, while its intensity decreases. The extent of peak broadening reflects the anion reactivity with the Au NP surface [265]. Besides, the spPV shows bell-shaped features revealing surface controlled processes. *E.g.* the reversible optically inferred acetate ion adsorption/desorption is correlated to redox processes identified on the electrochemical CV [29].

The first reports regarding the spectroelectrochemical inspection of EDL of a single Au nanorod [140,266], highlighted the opportunity for single NP scattering to probe the transfer of ~100 electrons, approaching the ultimate transfer of a single electron. It is however delicate to deconvolute the role played by the ion adsorption and Debye diffuse layer.

Even within the restricted evanescent layer optically probed by SPRM, it is still subtle deconvoluting the diffuse layer contribution from the charging of nanoobjects or that of the SPR electrode. If impedance-based SPRM is likely a way to address such contribution, yet it is operated under conditions of large electrode potential modulation. However, it allowed imaging EDL charging at metal nanowires [267] or graphene flakes [118]. When the optical signal is too low, the potential-modulation of the charge density is promisingly tracked [83,268] from the displacement of the scattering PSF centroid. The potential-modulated displacement enhances the tracking resolution to 0.1nm. It suggests a real time monitoring of the change in particle polarization associated to an inhomogeneous accumulation of electrons upon EDL charging.

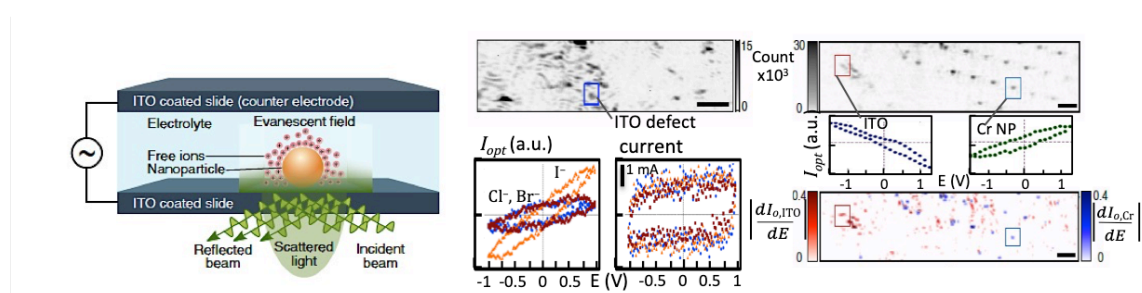

**Figure 23.** (a) TIR scattering imaging principle. (b) Image of the EDL at ITO nanodefects and potential modulation of their scattering signal in NaHalides together with the EC responses. (c) Enhancing the limit of detection of nanostructures, ITO nanodefects and Cr NPs have comparable scattering but opposite potential modulation, allowing a high-sensitivity potential-modulated reconstructed image. Adapted with permission from [269].

Ultra-sensitive optical detection indeed requires minimizing the background signal, by light confinement, *e.g.* offered by evanescent waves (Figure 23). Imaging the reflectance of the electrode-electrolyte interface by TIR enables such high sensitivity [69], further enhanced by the scattering/reflection detection mode analogous to that exploited in IRM. The image of an ITO-electrolyte interface reveals the scattering of some nanostructrured grains of the ITO surface [269]. The potential modulation of the scattering signal shows a higher sensitivity of these nanostructures to NaI (owing to the polarizability of $I^-$) compared to NaCl or NaBr electrolytes, not detected in the ensemble EC CV. The optical signal is sensitive to local EDL structuration. If a quantitative assignement is underway, the potential-modulated imaging enables further enhancement of the visualization discrimination of nanodefects: nanostructures vs single NPs (Figure 23).

## [6] Conclusion

Thanks to different scientific communities, optical microscopies have reached unprecedented level of sensitivity and resolution for the dynamic imaging of various physical, chemical and biological processes. While super-localization strategies enable imaging the displacement of various objects with nm resolution, advanced strategies and detection modes have allowed ultra-sensitive optical measurements down to the single molecule or single nanoobject level. There is now a wide range of detection modes which can be implemented to image as many experimental situations as possible, with an obvious prerequisite, that light may enter in. During the last two decades an increasing number of electrochemical situations have used an optical microscope as a means to see what is happening at an interface. The most used optical microscopies in electrochemistry have been presented.

More than seeing but to believe, electrochemical sciences need to evaluate a reaction yield. Optical microscopies are now mature enough to bring such quantitative analysis either from an event-counting strategy or from the analysis and understanding of the information gathered in an optical image. This has been described, through a first summary of the interaction of light with matter and for different concepts encountered in the literature (fluorescence, luminescence, Raman, scattering, surface plasmon resonance, refractive index and permittivity). As most optical microscopes are considered as linear instruments, a linear relationship exists between the observed optical signal and the "optical cross-section" of an object, meaning its refractive index or optical mass, or volume, absorbance, amount of emitter,… An effort has then been made to propose, at least in a first-approximation, generic expressions to exploit quantitatively optical images.

From the wide diversity of electrochemical situations coped by optical microscopy, the strategies employed to push optical imaging into the highest resolution and quantitative measurement have been reviewed, ranging from the micrometer to the nanometer scale, from ensemble to single entity responses. The fields of electrochemistry explored are vast, as solution species, adsorbates or solid material conversion can be deciphered.

If these studies have often considered model objects known for their large "optical cross-section" (response to an optical interrogation) some recent efforts have been made to explore a wider range of materials and electrochemical situations or reactions. More efforts are still needed in this direction and then to further improve the sensitivity of detection, for example to address, and benchmark, more systematically the materials used for energy conversion and storage, particularly supercapacitors and batteries electrodes. If they are more and more characterized by much more sophisticated instrumentation, at atomic resolution, optical microscopy could provide a much faster benchtop operando analysis.

The efforts made in the limit of detection are also considerable; for nanomaterials few examples convert or detect <$10^6$ atoms, approaching a level of complexity that can be computed by molecular dynamics.

Deploying optical and spectroscopic (hyperspectral) imaging techniques also means multiplying exponentially the amount of data and the time needed for their full analysis. Beyond the development of easy-to-use and open-source image analysis tools, one will need to recourse to image recognition and segmentation through machine learning and big data analysis. It is anyway a turning point to be taken now that electrochemical mechanisms rely on multi-correlative microscopies or characterization techniques.